\documentclass[draf]{agujournal2019}

\usepackage{amsmath}

\usepackage[inline]{trackchanges} %for better track changes. finalnew option will compile document with changes incorporated.
\usepackage{url} %this package should fix any errors with URLs in refs.
\usepackage{epstopdf,multirow,inputenc,ragged2e,subfigure,multirow,rotating,graphics,graphicx,txfonts,textcomp,epstopdf,arydshln}
\usepackage{float}
\usepackage{tabularray}

\usepackage{booktabs}
\usepackage[a4paper, total={6in, 8in}]{geometry}
\usepackage[markup=underlined]{changes}
\usepackage{array}
\usepackage{epstopdf}
\AppendGraphicsExtensions{.tif}
\usepackage{textcomp}

\usepackage{titlesec}
\usepackage{wrapfig}
\usepackage{circuitikz}
\usepackage{bm}
\usetikzlibrary{patterns, decorations.pathmorphing}

\tikzset{
    mybox/.style={
        font=\bfseries\LARGE,
        inner sep=5pt,
        text=black,
        align=center,
        line width=1.1pt,
        fill=lightgray, 
        draw=black,
    },
}

\usepackage{soul}

\definecolor{DarkGreen}{rgb}{0.2,0.5,0.2} % to color links in references

\makeatletter
\AtBeginDocument{\def\@citecolor{DarkGreen}}
\makeatother
\usepackage[nodisplayskipstretch]{setspace}
\usepackage{setspace}
\usepackage{soul}

\usepackage{colortbl}
\usepackage{apacite} 
\draftfalse

\journalname{Space Weather}

\begin{document}

%% ------------------------------------------------------------------------ %%
%
%  TITLE
%
%% ------------------------------------------------------------------------ %%

\title{Nowcasting Solar Energetic Particle Events for Mars Missions}

%% ------------------------------------------------------------------------ %%
%
%  AUTHORS AND AFFILIATIONS
%
%% ------------------------------------------------------------------------ %%

\authors{Jan Leo L\"owe\affil{1}, Salman Khaksarighiri\affil{1}, Robert F. Wimmer-Schweingruber\affil{1}, Donald M. Hassler\affil{2}, Bent Ehresmann\affil{2}, Jingnan Guo\affil{3, 4}, G\"unther Reitz\affil{5}
,Thomas Berger\affil{5}, Daniel Matthi\"a\affil{5}, Cary Zeitlin\affil{6}, Sven L\"offner\affil{1}}

\affiliation{1}{Institute of Experimental and Applied Physics, Christian-Albrechts-University, Kiel, Germany}
\affiliation{2}{Planetary Science Division, Southwest Research Institute, Boulder, CO, USA}
\affiliation{3}{Deep Space Exploration Laboratory/School of Earth and Space Sciences, University of Science and Technology of China, Hefei, PR China}
\affiliation{4}{CAS Center for Excellence in Comparative Planetology, USTC, Hefei, PR China}

\affiliation{5}{German Aerospace Center (DLR), Institute of Aerospace Medicine, Cologne, Germany}
\affiliation{6}{Leidos Corporation, Houston, TX, USA}

\correspondingauthor{Jan Leo Löwe, Salman Khaksarighiri and Robert F. Wimmer-Schweingruber}{loewe@physik.uni-kiel.de; khaksari@physik.uni-kiel.de; wimmer@physik.uni-kiel.de }

%% ------------------------------------------------------------------------ %%
%
%  KEYPOINTS
%
%% ------------------------------------------------------------------------ %%

%% Keypoints, final entry on title page.

%  List up to three key points (at least one is required)
%  Key Points summarize the main points and conclusions of the article
%  Each must be 100 characters or less with no special characters or punctuation and must be complete sentences

% Example:
% \begin{keypoints}
% \item	List up to three key points (at least one is required)
% \item	Key Points summarize the main points and conclusions of the article
% \item	Each must be 100 characters or less with no special characters or punctuation and must be complete sentences
% \end{keypoints}

\begin{keypoints}
\item We present a robust, easily operable, and reliable nowcasting system for SEP events, applicable in deep space and on the Martian surface.
\item Since its launch in November 2011, MSL/RAD has detected 5 SEP events route to Mars and 16 on its surface.
\item Our nowcasting system may provide astronauts with $>$30 minutes to avoid the peak and most of the radiation exposure from SEP events.
\end{keypoints}

%% ------------------------------------------------------------------------ %%
%
%  ABSTRACT + PLAIN LANGUAGE
%
%% ------------------------------------------------------------------------ %%

\begin{abstract}
\setlength{\parindent}{0pt} 
In addition to the omnipresent Galactic Cosmic Rays (GCRs), sudden solar energetic particle (SEP) events present considerable health hazards for manned space missions. These events not only contribute to an increased long-term cancer risk, but can, in extreme cases, cause acute radiation syndromes. Forecasting their imminent occurrence could significantly reduce radiation exposure by warning astronauts to move to shelter. However, all currently available tools are primarily designed for the Earth or Earth-Moon system, which limits their applicability to future Mars missions. To address this, we developed a nowcasting system for SEP events applicable in deep space and on the Martian surface, which serves as a reliable last-resort backup when forecasts fail. The methodology of this system is based on dose rates measured by the Radiation Assessment Detector (RAD) onboard the Mars Science Laboratory (MSL), which recorded 5 SEP events during the seven-month flight to Mars and 16 since its landing on Mars on August 6, 2012. An SEP event is triggered, and an astronaut is warned as soon as dose rates exceed the omnipresent background level by at least 25\%. This approach suggests that our system can provide astronauts with at least 30 minutes to avoid both peak radiation exposure and the majority of the cumulative dose from SEP events. Our nowcasting system is robust, easily implementable in real-life scenarios, and achieves a near-zero false alarm rate both in deep space and on the Martian surface.
\end{abstract}

\titlespacing*{\section}{0pt}{1.1\baselineskip}{\baselineskip}

\section*{Plain Language Summary}

\setlength{\parindent}{0pt} 

The on-site exploration and colonization of Mars is an ambitious future goal for humanity. However, with no natural shielding during flight, along with Mars' thin atmosphere and lack of a global magnetic field, astronauts on Mars missions are continuously vulnerable to persistent Galactic Cosmic Rays (GCRs) and sudden solar energetic particle (SEP) events. Both are very harmful to humans, increasing long-term cancer risk and causing severe effects on the central nervous system. Moreover, SEPs can potentially induce acute radiation syndromes, including nausea, vomiting, headache, fatigue, and fever. Predicting SEP events is therefore crucial to provide astronauts with sufficient time to seek shelter. However, all current predicting tools are specifically designed for Earth or Earth-Moon systems, and their applicability to Mars missions is not guaranteed. Accordingly, we developed a real-time warning system for Mars missions based on measurements from the Radiation Assessment Detector (RAD) onboard the Mars Science Laboratory (MSL), serving as a last-resort backup in cases where SEPs cannot be predicted. Our study suggests that the nowcasting system can provide astronauts with over 30 minutes to avoid most of the radiation exposure from SEP events. Our nowcasting procedure is easily feasible in real-life scenarios and can reliably detect SEP events.

\section{Motivation and Introduction}\label{sec_intro}

The deep-space radiation environment is primarily influenced by two sources due to their high energy and prolonged exposure: sporadic eruptive events consisting of Solar Energetic Particles (SEPs) and the omnipresent Galactic Cosmic Rays (GCRs). The latter originate outside the solar system, primarily from supernovae and neutron star mergers (for nuclei with Z $\gtrdot$ 30) \cite{doi:10.1126/science.aaq0049}. GCRs typically exhibit energies ranging from less than 1 MeV/nuc to hundreds of TeV and are predominantly composed of protons and helium ions, which constitute approximately 87\% and 12\% of GCR nuclei, respectively. Additionally, 1-2\% heavier nuclei with charges ranging from Z=3 (lithium) to about Z=26 (iron) are also present in GCRs \cite{simpson1983, Mewaldt1994}. Upon entering the solar system, GCR flux and energy spectrum variations are influenced by the Sun's varying activity and the associated changes in the solar wind, as well as the interplanetary magnetic field \cite{https://doi.org/10.1029/2010JA016105, Badhwar1994}. This phenomenon, designated solar modulation, is correlated with the 11-year solar activity cycle and the approximately 27-day heliospheric rotation, which particularly shields low-energy GCRs from penetrating into the heliosphere \cite{Wiedenbeck2007, gieseler2018, simpson1983}. Fluctuations in the intensity of GCRs can also result from Forbush decreases, which are characterized by sudden reductions in the intensity of GCRs followed by gradual recoveries. These events are typically observed during the passage of coronal mass ejections (CMEs) through interplanetary space \cite{refId0, Dumbovic_2018}. The second source, SEP events, are sporadic, yet highly variable in intensity, consisting of energetic particles accelerated in proximity to the Sun \cite{Reames1999, Desai2016}. These events are more frequent during solar maximum. SEPs mainly comprise protons, but also include electrons and heavier ions, with energies ranging from suprathermal levels of a few keV to several hundreds of MeV, and occasionally up to a few GeV \cite{guo2021radiation}. At these energies, SEP events can reach significantly higher fluxes compared to GCRs.

Upon reaching Mars, GCRs and SEPs interact with the Martian atmosphere, lose energy through ionization processes, and generate a cascade of secondary particles through nuclear reactions such as spallation and fragmentation \cite{saganti2004radiation, feldman2002global}. These interactions are closely linked to the variability of the Martian atmosphere, which undergoes diurnal and seasonal changes \cite{rafkin2014diurnal, guo2015modeling}. Diurnal pressure fluctuations are caused by thermal tides, leading to oscillations in column mass between day and night. Seasonal variations in surface pressure are governed by the cycling of the cold and warm CO$_2$ polar ice caps \cite{https://doi.org/10.1002/2016JE005206, Gomez-Elvira2012}. Due to Mars' atmosphere being significantly thinner than Earth's, with an average surface atmospheric density of $\rho_{\mathrm{Mars}}\sim0.020\frac{\mathrm{kg}}{\mathrm{m}^3}$ ($\sim$ 1\% of Earth's) \cite{williams2024mars}, both secondary particles as well as primary GCRs and SEPs can penetrate it and reach the surface. For primary particles to reach the Martian surface, they must exceed the atmospheric cutoff energy, which for protons is $\sim150$ MeV \cite{Guo_2019_Cutoff, khaksarighiri2023zenith, https://doi.org/10.1029/2021JE007157}. These surviving particles interact with regolith nuclei, creating an upward 'albedo' radiation field composed of charged particles and also neutrons and gamma rays \cite{appel2018detecting, https://doi.org/10.1029/2022SW003344, https://doi.org/10.1029/2021GL093912}. A recent detailed overview of the Martian radiation environment based on both modeling and observational results can be found in \cite{2021A&ARv..29....8G}.

According to the International Commission on Radiological Protection (ICRP) Publication 103, the recommended annual dose limit for public exposure is 1 mSv, while for occupational exposure it is 20 mSv \cite{ICRP2007_new}. For astronauts, the total career dose limits due to space radiation exposure must be less than 600 mSv, regardless of age and gender \cite{SHAVERS202414}. Due to the absence of natural shielding in deep space, combined with the Mars' thin atmosphere and the lack of an intrinsic global magnetic field, astronauts and spacecraft systems on Mars missions are continuously exposed to elevated levels of radiation, which can pose significant health risks to astronauts. A distinction is made between acute radiation syndroms (ARS), primarily triggered by SEP events, and long-term effects, caused by the combination of GCRs and SEPs \cite{Barcellos-Hoff2015}. The latter particularly increases the risk of cancer through carcinogenesis \cite{Cucinotta2006} and may potentially lead to degenerative tissue defects \cite{Boerma2015, Baselet2016, Smart2017}. Impairments to the central nervous system may also affect human health, including reduced motor function, changes in behavior, and neurological disorders \cite{Cucinotta2014, khaksarighiri2021easy}. In addition, cosmic radiation can damage deoxyribonucleic acid in both cultured cells and blood cells of astronauts, leading to mutations and the generation of reactive oxygen species in vivo \cite{Moreno-Villanueva2017}. The omnipresent radiation exposure during a $\sim$500-day round trip to Mars results in an estimated dose equivalent of approximately 930 mSv, significantly exceeding both the annual ICRP exposure limits and the career dose limits for astronauts \cite{https://doi.org/10.1029/2019SW002354, doi:10.1126/science.1235989}. This estimated dose does not even include the additional radiation exposure during the stay on Mars. ARS include symptoms such as nausea, diarrhea, headache, and anorexia \cite{inbook, Cucinotta2014}. Fatigue, fever, and skin injuries may also be triggered \cite{2021hbba.book..263M}. Furthermore, \cite{10.3389/fspas.2023.1117811} demonstrated that acute exposures to SEP protons and iron ions disrupt intestinal function, potentially inducing mucosal lesions, vascular congestion, epithelial barrier breakdown, and significant enlargement of mucosa-associated lymphoid tissue. SEP events, such as the August 1972 event, can even be so powerful that the dose experienced by space-suited astronauts could lead to lethal consequences \cite{tranquille1994solar, HandWiki2024, Parsons2000}. 
 
Mitigating radiation exposure for astronauts is therefore essential for Mars missions. Strategies adhering to the ALARA (as low as reasonably achievable) principle leverage Martian topography, such as constructing bases near buttes \cite{ehresmann2021natural} or on water-rich subsurfaces \cite{2020JGRE..12506246R}. Other approaches focus on optimizing shielding materials in spacecraft and spacesuits to reduce radiation exposure to further minimize radiation risks \cite{https://doi.org/10.1029/2021SW002749, Montesinos2021, khaksarighiri2020calculation, GOHEL2022110131}. However, protecting against GCRs is generally challenging due to the high kinetic energy of the particles, which results in strong penetration capabilities and a significant probability of generating secondary particles in shielding materials (e.g., spacecraft, spacesuits, or the Martian atmosphere). SEP events, on the other hand, primarily consist of lower-energy particles. Consequently, effective shielding should be sufficient to protect astronauts. The straightforward solution is to seek refuge in specially designed spacesuits during flight and in either dedicated shelters or underground bunkers on the Martian surface. 

However, the challenge lies in the unpredictability of SEP events. Astronauts, particularly during extravehicular activities (EVA) or surface operations, require time to reach shelter. To address this, forecasting tools such as REleASE \cite{RELEASE}, ESPERTA \cite{Laurenza_2018}, and COMESEP \cite{COMESEP} have been developed to predict the timing, intensity, and trajectory of SEP events.  Nevertheless, forecasting accuracy remains limited, typically ranging between 65\% and 80\%, depending on the model employed \cite{WHITMAN20235161}. This limitation can be attributed to two key factors: First, the dynamics of active regions on the Sun and the generation of flares remain poorly understood. Second, data and instrumentation are limited, as is the complexity of the prediction models. Moreover, the currently available forecasting systems are specifically designed for the Earth or Earth-Moon system and cannot be directly applied to an entire Mars mission. Therefore, we develop a straightforward and reliable nowcasting system for Mars missions, which acts as a fallback option when forecasts are unavailable.

In Section \ref{Chapter_RAD}, we begin by describing the detector and the necessary datasets used in our study. Section \ref{Chapter_Methodology} outlines the methodology of our nowcasting system and the information that must be provided to an astronaut. Subsequently, in Section \ref{Chapter_Result}, we present our results. We evaluate how long an astronaut would have needed to stay in a shelter, how much radiation exposure could have been avoided, and whether our nowcasting system could have provided an astronaut with enough time to seek shelter. In Subsection \ref{Chapter_False_Alarm}, we study the reliability of our nowcasting system. Finally, we summarize our findings in Section \ref{Chapter_Summary}.

\section{The Radiation Assessment Detector}\label{Chapter_RAD}

\begin{wrapfigure}{r}{0.5\textwidth}
\centering
\resizebox{0.255\textwidth}{!}{%
\begin{circuitikz}
\tikzstyle{every node}=[font=\Large]
%Detector A
\draw [fill=blue] (7.5,13.5) rectangle (12.5,13.75);
\draw [fill=cyan] (8.75,13.75) rectangle (11.25,13.5);
\draw [line width=0.7pt, ->, >=Stealth] (6,20.875) -- (7.25,20.875);
\node [font=\huge] at (5.5,20.875) {\textbf{A}};
%Detector B
\draw [ fill=blue] (7.5,13.75) rectangle (12.5,14);
\draw [fill=cyan] (8.75,14) rectangle (11.25,13.75);
\draw [line width=0.7pt, ->, >=Stealth] (6,13.875) -- (7.25,13.875);
\node [font=\huge] at (5.5,13.875) {\textbf{B}};
%Detector C
\draw [fill=blue] (7.5,21) rectangle (12.5,20.75);
\draw [fill=cyan] (8.75,21) rectangle (11.25,20.75);
\draw [line width=0.7pt, ->, >=Stealth] (14,13.65) -- (12.75,13.65);
\node [font=\huge] at (14.5,13.65) {\textbf{C}};
%Detector D
\draw [fill=gray] (8.75,13.5) -- (11.25,13.5) -- (12.5,11) -- (7.5,11) -- cycle;
\node [font=\huge] at (10,12.25) {\textbf{D}};
%Detector E
\draw [fill=red] (7.5,10.75) rectangle (12.5,8.5);
\node [font=\huge] at (10,9.5) {\textbf{E}};
%Detector F1
\draw [fill=yellow] (6,8) -- (7.25,8) -- (7.25,11) -- (8.5,13.5) -- (7.25,13.5) -- (6,11) -- cycle ;
\node [font=\huge] at (6.6,9.5) {\textbf{F1}};
\draw [fill=yellow]  (12.75,8) -- (14,8) -- (14,11) -- (12.75,13.5) -- (11.5,13.5) -- (12.75,11) -- cycle;
\node [font=\huge] at (13.3,9.5) {\textbf{F1}};
%Detector F2
\draw [fill=yellow] (6,8) rectangle (14,6.75);
\node [font=\huge] at (10,7.35) {\textbf{F2}};
%Cone
\draw [line width=0.5pt] (7.5,21) rectangle (12.5,13.5);
%Viewcones
\draw [line width=1pt, dashed] (11,14) -- (6.25,23.5);
\draw [line width=1pt, dashed] (9,14) -- (13.75,23.5);
\draw[<->, thick, bend left] (6.25,23.5) to (13.75,23.5);
\node [font=\huge] at (10,24.1) {\textbf{A$\times$B}};
\draw [line width=1pt, dotted] (11.15,14) -- (8.25,22.25);
\draw [line width=1pt, dotted] (8.75,14) -- (11.75,22.25);
\draw[<->, thick, bend left] (8.25,22.25) to (11.75,22.25);
\node [font=\huge] at (10,22.25) {\textbf{A2$\times$B}};

\end{circuitikz}
}
\caption{Schematic view of RAD. RSH houses three silicon detectors (A, B, C) alongside a caesium iodide scintillator (D) and a plastic scintillator (E). Both scintillators are encased in a plastic anticoincidence shield (F1, F2). The two different FOVs for charge particle and LET measurements are A2$\times$B = 36° and A$\times$B = 60°.}
\label{fig:RAD}
\end{wrapfigure}
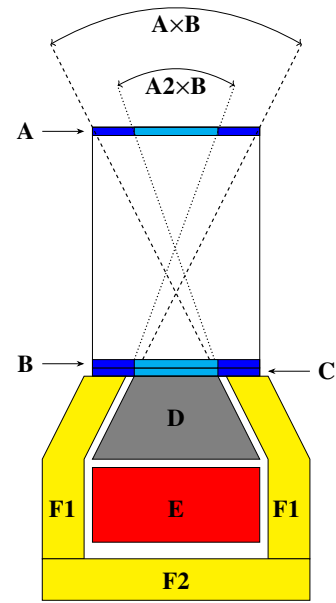

The Radiation Assessment Detector (RAD; \cite{hassler2012radiation}) is installed on the Mars Science Laboratory (MSL; \cite{Grotzinger2012}) rover 'Curiosity'. The RAD sensor head (RSH) integrates multiple detectors to effectively measure energetic particles. It includes three thin hexagonal silicon (Si) diodes (A, B, C) at the top of the detector stack, each with a thickness of 300 $\mu$m. Below these are a thallium-doped cesium iodide (CsI(Tl)) scintillator (D), shaped as a hexagonal pyramid and measuring 28 mm in height, along with a plastic scintillator (E) with a thickness of 18 mm. The Si detectors are subdivided into inner and annular outer segments with separate readouts, with segments of detector A designated as A2 (inner) and A1 (outer). Scintillators D and E are encapsulated within an additional plastic scintillator (F) used to implement anticoincidence logic. Detector F is divided into areas surrounding D and E (F1) and beneath E (F2). A schematic representation of the RSH is depicted in Figure \ref{fig:RAD}.

RAD started operating on December 6, 2011, measuring the radiation environment in deep space during the 253-day cruise to Mars. Since landing on August 6, 2012, RAD has been monitoring the radiation environment on the Martian surface. 

In addition to detecting charged particles (cf. \cite{https://doi.org/10.1002/2013JE004547}) and neutral particles (cf. \cite{https://doi.org/10.1002/2013JE004539, GUO201712}), RAD also measures dosimetric quantities such as the total absorbed ionizing dose rate and Linear Energy Transfer (LET) spectra. LET quantifies the amount of energy transferred to a material by ionizing radiation per unit length d$E$/d$l$ \cite{harrison2021icrp, cucinotta2003radiation}. Detectors A and B are coincident to define the acceptance angle for the LET spectra, as detailed in \cite{Zeitlin2019_LET}: the inner field of view (FOV) is A2$\times$B = 36°, and the outer FOV is A$\times$B = 60°. Absorbed dose, expressed in Gray (Gy), is defined as the average energy deposited by ionizing radiation per unit mass d$m$. RAD measures the dose rates in $\mu$Gy/day with a full 4$\pi$ steradian acceptance angle using silicon detector B and scintillator E. The former is labeled 'dose rate B' and the latter 'dose rate E'. Since the scintillator E has a composition similar to human tissue, the dose rate E represents a tissue-equivalent dose rate. Alongside the 15-minute dose rate data products, RAD also provides high-resolution dose rate measurements as one-minute data products. 

\section{Methodology} \label{Chapter_Methodology}

The primary objective of a nowcasting system is to provide astronauts with the following critical information:

\begin{itemize} 
    \item[1.] When should an astronaut seek shelter? 
    \item[2.] When is it safe for an astronaut to leave the shelter? 
\end{itemize}

This requires a methodology that is not only reliable but also optimized for the rapid identification of SEP events. Our nowcasting system is based on the dose rate E measured by MSL/RAD, which has proven to be the most suitable parameter for nowcasting, could be the most easily accessible real-time data, and enables a direct assessment of the severity of an SEP event. Although dose rate B can also be used for the analysis of SEP events, it is not suitable for nowcasting, as discussed in the Appendix \ref{SM_NowcastingB}.

Figure \ref{fig_DoseE} provides an overview of the dose rate E measured in $\sim$15-minute cadences by MSL/RAD, spanning the mission period from launch on November 26, 2011, to September 9, 2024. The left panel shows dose rate E during the seven-month flight to Mars, while the right panel presents measurements obtained on the Martian surface since August 6, 2012. The average dose rate E in deep space is (443.14$\pm$21.25)$\mu$Gy/day, approximately twice as high as on the Martian surface (248.01$\pm$48.79)$\mu$Gy/day, due to Mars' planetary shielding. Various factors contribute to dose rate E, including Forbush decreases, solar rotation, Martian atmospheric variations, and topographic shielding, as discussed in Section \ref{sec_intro}. These effects, however, can only be inferred on these long scales. More apparent is the anti-correlation between dose rate E and solar activity, as analyzed in \citeA{2021A&ARv..29....8G}. The mission began during the solar maximum of Solar Cycle 24, reached the solar minimum around 2020, and is currently approaching the solar maximum of Solar Cycle 25. Detected SEP events are marked with red arrows, showing a higher frequency during solar maximum. In total, 5 SEP events were recorded in deep space during the transit to Mars and 16 were measured on the Martian surface. Most of these events significantly exceed the omnipresent radiation background within a short duration. The largest event measured during the cruise to Mars occurred on March 7, 2012, while the most intense event ever recorded on the surface of Mars took place on May 19, 2024.

\begin{figure}[H]
    \centering
    \includegraphics[width=0.95\textwidth]{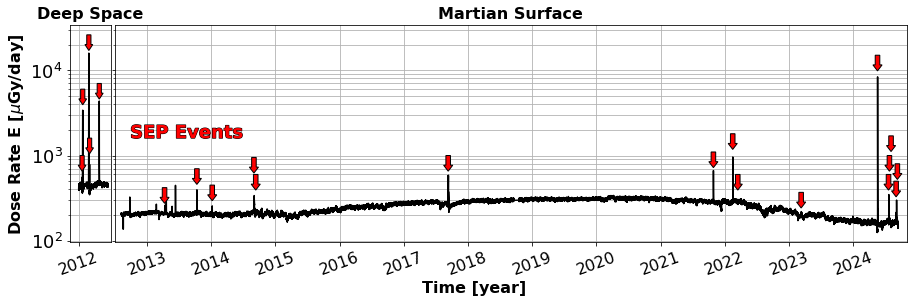}
    \caption{Dose rate measured by the MSL/RAD plastic scintillator detector E in deep space since the launch on November 26, 2011, during the seven-month flight to Mars (left) and on the Martian surface from landing on August 6, 2012, to September 9, 2024 (right). Detected SEP events are indicated by red arrows.}
    \label{fig_DoseE}
\end{figure}

\begin{figure}[H]
    \centering
    \includegraphics[width=1.0\textwidth]{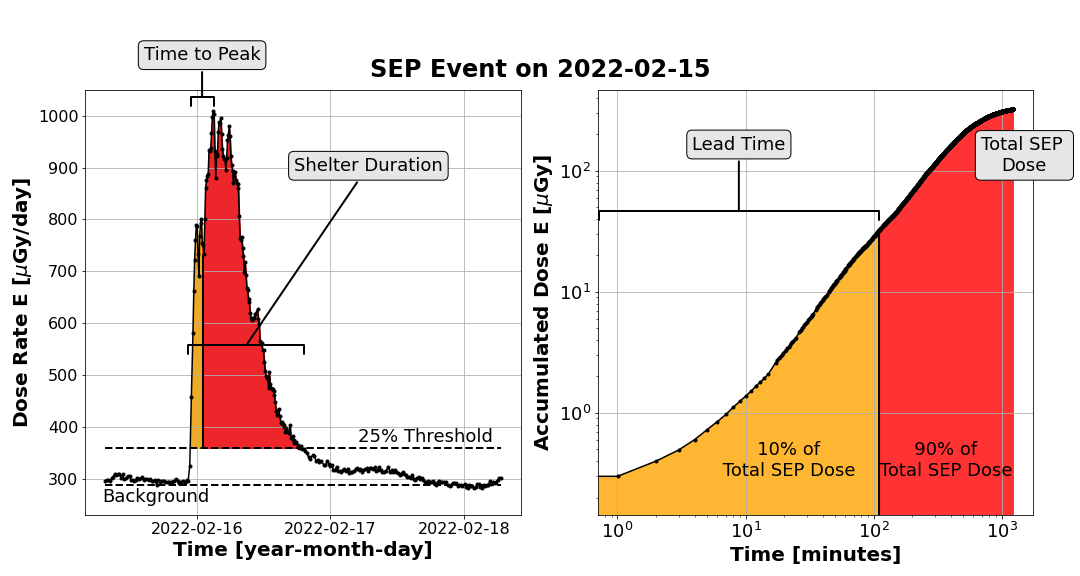}
    \caption{Definitions of the parameters that a nowcasting system must provide an astronaut, exemplified by the SEP event observed by MSL/RAD on the surface of Mars on February 15, 2022. The left panel shows the dose rate E as a function of time, with the fitted background and the 25\% trigger threshold marked as horizontal dashed lines. The color-shaded area indicates the SEP event. The shelter duration is the time from onset to end of the SEP event, and the time to peak is the time from onset to the maximum of dose rate E. On the right, the background-removed accumulated dose E per minute between onset and end of the SEP event is shown, resulting in the total SEP dose. The lead time is defined as the period in which an astronaut can avoid at least 90\% of the total SEP dose (time region marked in red) and would receive only 10\% of the total SEP dose (time region marked in orange).}
    \label{fig_Definitions}
\end{figure}

To establish a consistent definition of when astronauts need to go to shelter, it is first necessary to characterize the omnipresent radiation background. We apply a constant linear fit to the dose rate E measured in 15-minute cadences over the five days preceding the SEP event. The background fits are stored in Appendix\ref{SM_Background}. This approach helps to smooth out daily variations caused by atmospheric pressure changes, topography-induced rover inclination shifts, and minor solar influences. Seasonal changes in atmospheric column depth, solar modulation effects, and the Sun's $\sim$26-day heliospheric rotations are considered negligible over a five-day period. An SEP event is considered triggered when the dose rate exceeds the background level by more than 25\%, corresponding to a threshold of 1.25 times the background dose rate. An astronaut must enter a shelter as soon as possible (ASAP) once our nowcasting system is triggered. It should be emphasized that this approach is not intended to determine the exact physical onset of an SEP event, but rather to identify an SEP event as quickly, effectively, and reliably, while ensuring that the system is practical for real-life implementation. Whether the 25\% trigger threshold meets these criteria is discussed in Subsection \ref{Chapter_False_Alarm}. The end of an SEP event is defined when the dose rate E drops below the 25\% trigger threshold. Our nowcasting system then informs the astronaut that it is safe to leave the shelter. The procedure exemplified for the SEP event on February 15, 2022, is depicted in Figure \ref{fig_Definitions} (left). The fitted background and the resulting 25\% trigger threshold are marked with horizontal dashed lines. The SEP event, according to our definitions of triggering and end, is highlighted as a colored area (orange and red combined).

The dataset illustrated in Figure \ref{fig_DoseE} is used to test our nowcasting methodology and to develop suggestions for future protective measures for astronauts based on historical SEP events. In particular, the following questions will be addressed:

\begin{itemize}
    \item[3.] How long would an astronaut need to stay in the shelter? 
    \item[4.] How much radiation exposure could have been avoided? 
    \item[5.] Could our nowcasting system still have provided an astronaut with sufficient time to take shelter? 
\end{itemize}

To address question 3, we assume the best-case scenario in which an astronaut is already in a shelter at the onset of the SEP event and remains there for its entire duration. Therefore, the 'shelter duration' is the time calculated from onset to end of the SEP event. For this time period, we calculate the background-removed accumulated SEP dose, referred to as the 'total SEP dose' (see Figure \ref{fig_Definitions}, right). This corresponds to the radiation exposure discussed in question 4, which could have been avoided in the best-case scenario. Nevertheless, this dose represents a conservative upper limit, as the shielding for an astronaut in deep space and on the Martian surface differs from that used in the measurements presented here.

Of course, an astronaut should seek shelter ASAP once an SEP event triggers our nowcasting system. However, especially during EVAs or operations on the surface of Mars, astronauts require time to reach safety. The optimal scenarios considered in questions 3 and 4 are therefore not always guaranteed. Consequently, in question 5, we aim to determine whether our nowcasting system could have provided an astronaut with enough time to seek shelter and whether this could help establish limits on how far, in terms of time, an astronaut is allowed to move away from a shelter during future Mars missions. Given the current lack of a unified definition for the time needed to reach safety once an SEP event has already impacted an astronaut, we investigate two different scenarios. First, we analyze the 'lead time', defined as the time an astronaut has to avoid at least 90\% of the total SEP dose, thus mitigating the primary portion of radiation exposure (see Figure \ref{fig_Definitions}, right). Second, we examine the 'time to peak', which represents the time available to an astronaut to avoid the peak dose rate of an SEP event (see Figure \ref{fig_Definitions}, left). The former definition may be particularly relevant for mitigating additional doses that contribute to long-term radiation risks, while the latter could be crucial for preventing ARS.

The final crucial question that must be addressed for a functional nowcasting system is:

\begin{itemize}
\item[6.] Can the nowcasting system reliably detect SEP events? 
\end{itemize}

This means that, to be implementable for Mars missions, the nowcasting system should be triggered exclusively by SEP events in deep space and on the Martian surface, while remaining unresponsive to SEP-unrelated outliers in the measurements. To answer the question 6, we simulate a real-life scenario using the dataset from Figure \ref{fig_DoseE}, as described and analyzed in Subsection \ref{Chapter_False_Alarm}.

\section{Results and Discussion}\label{Chapter_Result}

Following the methodology described in Section \ref{Chapter_Methodology}, we analyzed all SEP events concerning the introduced shelter duration, time to peak, lead time, and total SEP dose. The results, categorized into deep space and on the Martian surface, are stored for each respective SEP event in Table \ref{tab:Results}. 

\begin{table}[H]
    \centering
    \renewcommand{\arraystretch}{1.5} 
    \caption{Analysis of shelter duration, time to peak, lead time, and total SEP dose for the SEP events measured with MSL/RAD. The events are categorized into those measured in deep space and on the surface of Mars. The grey-highlighted rows correspond to the largest SEP event measured in deep space and on the surface of Mars, respectively.}
    \begin{tabular}{@{\extracolsep{0pt}}
    >{\centering\arraybackslash}p{0.1cm}>{\centering\arraybackslash}p{2.78cm} | >{\centering\arraybackslash}p{3.03cm} | >{\centering\arraybackslash}p{3.03cm} >{\centering\arraybackslash}p{3.03cm} | >{\centering\arraybackslash}p{2.32cm}
    }
    \toprule
     & \textbf{SEP Event [year-month-day]} & \textbf{Shelter Duration [day\space\space hour:minute]} & \textbf{Time to Peak \space\space\space\space[day\space\space  hour:minute]} & \textbf{Lead Time\space\space\space\space\space\space\space\space\space  [day\space\space hour:minute]} & \textbf{Total SEP Dose [$\mu$Gy]}\\
    \midrule
    \parbox[t]{1mm}{\multirow{5}{*}{\rotatebox[origin=c]{90}{\textbf{DEEP SPACE}}}} 
    & 2012-01-23 & 00d\space\space03h:09m & 00d\space\space02h:09m & 00d\space\space00h:17m & 12.71   \\
    & 2012-01-27 & 02d\space\space05h:54m & 00d\space\space02h:52m & 00d\space\space02h:11m & 1466.30 \\
    & \cellcolor{gray!20}2012-03-07 & \cellcolor{gray!20}03d\space\space07h:33m & \cellcolor{gray!20}01d\space\space12h:34m & \cellcolor{gray!20}00d\space\space16h:43m & \cellcolor{gray!20}9901.40 \\
    & 2012-03-13 & 00d\space\space10h:16m & 00d\space\space00h:36m & 00d\space\space00h:35m & 86.05   \\
    & 2012-05-17 & 01d\space\space00h:43m & 00d\space\space01h:25m & 00d\space\space00h:51m & 1170.20 \\
    \midrule
    \parbox[t]{1mm}{\multirow{16}{*}{\rotatebox[origin=c]{90}{\textbf{MARTIAN SURFACE}}}} 
    & 2013-04-10 & 00d\space\space05h:33m & 00d\space\space01h:28m & 00d\space\space01h:28m & 9.22   \\ 
    & 2013-10-10 & 00d\space\space11h:15m & 00d\space\space02h:07m & 00d\space\space01h:07m & 32.32   \\
    & 2014-01-05 &           &           &           &         \\
    & 2014-09-02 & 01d\space\space14h:17m & 00d\space\space09h:01m & 00d\space\space02h:47m & 125.93  \\
    & 2014-09-10 & 00d\space\space01h:41m & 00d\space\space01h:02m & 00d\space\space00h:02m & 2.09    \\
    & 2017-09-10 & 01d\space\space08h:01m & 00d\space\space06h:49m & 00d\space\space03h:44m & 261.79  \\
    & 2021-10-28 & 00d\space\space17h:47m & 00d\space\space05h:12m & 00d\space\space02h:01m & 161.44  \\
    & 2022-02-15 & 00d\space\space20h:31m & 00d\space\space03h:55m & 00d\space\space01h:50m & 323.75  \\
    & 2022-03-14 &           &           &           &         \\
    & 2023-03-12 & 00d\space\space01h:23m & 00d\space\space01h:08m & 00d\space\space00h:05m & 1.22    \\
    & \cellcolor{gray!20}2024-05-20 & \cellcolor{gray!20}01d\space\space17h:55m & \cellcolor{gray!20}00d\space\space00h:53m & \cellcolor{gray!20}00d\space\space00h:57m & \cellcolor{gray!20}1730.55 \\
    & 2024-07-22 & 00d\space\space13h:16m & 00d\space\space02h:16m & 00d\space\space01h:20m & 57.98   \\
    & 2024-07-26 &           &           &           &         \\
    & 2024-08-04 &           &           &           &         \\
    & 2024-09-02 & 00d\space\space01h:25m & 00d\space\space00h:34m & 00d\space\space00h:08m & 2.89    \\
    & 2024-09-05 & 00d\space\space08h:35m & 00d\space\space00h:33m & 00d\space\space00h:33m & 25.52   \\
    \bottomrule
    \end{tabular}
    \label{tab:Results}
\end{table}

Four SEP events on the surface of Mars do not exceed the 25\% trigger threshold and therefore cannot be identified using our methodology. However, these events are not relevant in the context of nowcasting, as they do not pose an additional threat to astronauts. Since they can still be identified through other post-processing methods, such as studying proton flux as described by \cite{https://doi.org/10.1029/2018GL077801}, or comparing MSL/RAD measurements with those from magnetically connected spacecraft, cf. \cite{https://doi.org/10.1029/2023GL103069}, their timestamps are included here to ensure complete documentation of all SEP events measured with MSL/RAD.

Additionally, some events only slightly exceed our 25\% threshold, specifically the event on January 23, 2012 in deep space, and the events on the Martian surface on September 10, 2014, March 12, 2023, and September 2, 2024. These events are only above the GCR background for a very short period, with shelter duration varying between approximately 1 hour and 20 minutes to about 3 hours. Consequently, the time to peak can also be very short, ranging from about 30 minutes to approximately 2 hours. The lead time can be even shorter, ranging from a few minutes to up to an hour. However, these SEP events also result in negligible additional radiation exposure above the GCR background, as indicated in the total SEP dose in Table \ref{tab:Results}. In particular, since shielding for astronauts will be more effective than in the presented measurements, as mentioned in Section \ref{Chapter_Methodology}. Therefore, these SEP events are also not prioritized events that require specific avoidance.

Considering the remaining SEP events, the shelter duration varies between at least about 5 hours and 30 minutes to several days, with a maximum of about 3 days and 7 hours. The average shelter duration in deep space of 01d 18h:06m compared to on the surface of Mars of 00d 21h:01m suggests that the shelter duration in deep space is longer than on the surface of Mars. This may be attributed to the fact that late arrivals of low-energy particles still contribute to the shelter duration in deep space, whereas they are shielded by the atmosphere on Mars and therefore do not affect the shelter duration. The average time to peak in deep space is 00d 10h:21m and on the surface of Mars 00d 03h:37m. Findings from SEP events observed during the STEREO and Helios era (cf. \cite{richardson289labrador, kallenrode1993neutral}) indicate that the time to peak significantly depends on how well the SEP event source is magnetically connected to the observer, in this case, the transfer vehicle or Mars \cite{https://doi.org/10.1029/2019SW002354}. The dose rate E from SEP protons that are not well magnetically connected tends to increase more slowly compared to well-connected events. The average lead time in deep space is 00d 05h:05m and on the Martian surface 00d 01h:38m. Similarly to shelter duration, both the time to peak and the lead time tend to be longer in deep space than on the surface of Mars. Furthermore, it can be observed that time to peak $\gtrsim$ lead time. To propose a universal lower limit for the time an astronaut has to reach shelter, we therefore assume the minimum lead time, which is approximately 30 minutes. In other words:

\begin{quote}
\normalsize Our study suggests that, using our nowcasting system, an astronaut should not move more than 30 minutes away from a shelter in order to still avoid not only more than 90\% of the total SEP dose but also the peak dose rate in the case of an SEP event.
\end{quote}

In Table \ref{tab:Results}, the largest SEP events measured in deep space and on the surface of Mars are highlighted in grey. For the deep space event on March 7, 2012, an astronaut would have had approximately one and a half days using our nowcasting system to avoid the peak dose rate E and more than 16 hours to avoid at least 90\% of the total SEP dose. The latter corresponds to a mitigated dose of approximately 8.9 mGy, which is almost 9 times the annual ICRP dose limit for public exposure and $\sim$44.5\% of the annual ICRP dose limit for occupational exposure. For the largest SEP event ever measured on the surface of Mars on May 20, 2024, there would be just over 50 minutes to avoid both the peak dose rate and 90\% of the total SEP dose. This corresponds to a mitigated dose of approximately 1.56 mGy, which is almost 1.6 times the annual ICRP dose limit for public exposure and $\sim$15.6\% of the annual ICRP dose limit for occupational exposure. Both cases exemplify the critical importance of minimizing radiation exposure from SEP events.

\subsection{False Alarm Rate}\label{Chapter_False_Alarm}

\begin{figure}[H]
    \centering
    \includegraphics[width=0.9\textwidth]{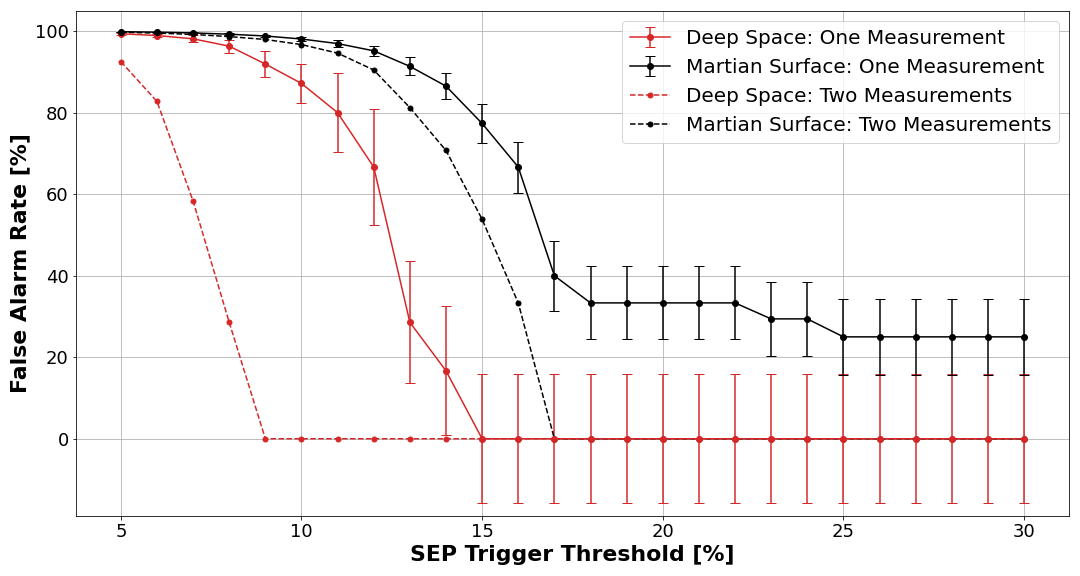}
    \caption{False alarm rates in deep space (red) and on the Martian surface (black) for our nowcasting system, depending on different SEP trigger thresholds. Solid lines represent warnings triggered as soon as a single 15-minute dose rate E measurement exceeds the respective threshold. Dashed lines represent the FAR when two consecutive dose rate E measurements exceed the respective trigger threshold. The statistical errors for the dashed lines are analog to those for the one-measurement results and are therefore not shown for clarity.}
    \label{fig_FalseAlarm}
\end{figure}

So far, we have examined the SEP events detected by MSL/RAD and demonstrated the potential of a nowcasting system for Mars missions. In this subsection, we address question 6, testing whether our nowcasting system can reliably detect SEP events in deep space and on the Martian surface. For this purpose, we use the 15-minute dataset presented in Figure \ref{fig_DoseE} to simulate a real-life scenario. We apply the methodology introduced in Section \ref{Chapter_Methodology}, thus fitting the background over the last 5 days and comparing the measurements of the current day against it. If the measurements of the current day exceed a certain trigger threshold (set at $>25\%$ in this paper), this is considered as triggering an SEP event. Using this approach, we recursively loop through the dataset, incrementing by one day each iteration. Subsequently, we check whether an SEP event actually occurred when our nowcasting system was triggered or if it was erroneously triggered, meaning no SEP event occurred. The latter can happen, e.g., if rare heavy ions from GCRs contribute to the dose rate E, or when the dose rate E returns to the original background after a Forbush decrease. We calculate the false alarm rate (FAR) as follows:

\begin{equation} 
    \text{FAR} = \frac{\text{false trigger}}{\text{false trigger} + \text{correct trigger}} 
\end{equation}

which is an indicator of the probability of a false identification of the SEP event. The FAR should be minimized to ensure effective nowcasting. In addition to the 25\% above background threshold used in this study, we also test trigger thresholds ranging from 5\% to 30\% above background to determine the optimal threshold for nowcasting. The results are plotted in Figure \ref{fig_FalseAlarm}. The error bars represent the statistical uncertainty, assuming a binomial distribution of either correct or incorrect triggering. We distinguish between the FAR in deep space (red) and on the surface of Mars (black). In addition, we examine two different cases. In the first, we consider triggering an SEP event as soon as a single dose rate E measurement exceeds the respective threshold. In a real-life implementation, this would immediately alert the astronaut. In the second case, two consecutive measurements must exceed the threshold. The second measurement is used as a cross-check to either verify or falsify the SEP event. In a real-life implementation, the astronaut is first notified as soon as the nowcasting system has been triggered by a single measurement above the threshold, allowing them to start preparing for potential sheltering. Only once the second measurement confirms the SEP event,the astronaut will receive an official alert.The advantage of the second case over the first is its reduced sensitivity to background dose outliers unrelated to SEP events. However, the drawback is the sacrifice of valuable warning time (15 minutes), resulting in shorter lead time and time to peak for astronauts. In Figure \ref{fig_FalseAlarm}, case 1 is shown with solid lines, and case 2 with dashed lines. 

The FAR in deep space for case 1 drops to 0\% with a trigger threshold of 15\% above the background, while in case 2, this already occurs with a trigger threshold of 9\%. On the surface of Mars, the FAR based on two measurements is 0\% for trigger thresholds of $\gtrsim$17\%. The FAR on the Martian surface in case 1 remains nearly constant at $\sim$33\% for a trigger threshold of $\gtrsim$18\%, but drops to 25\% at a trigger threshold of 25\% above the background.

For our nowcasting system, we have assumed a uniform trigger threshold of 25\% in deep space and on the Martian surface. At this threshold, our nowcasting system achieves a zero FAR in both deep space and on the Martian surface under the cross-checking assumption. For direct warning based on a single measurement, the FAR is zero in deep space and 25\% on the Martian surface. This is considered in the paper as the best balance between high-precision nowcasting with a uniform trigger threshold in deep space and on the Martian surface while minimizing the risk of excessive radiation exposure. In real-life implementation, it could even be suggested to lower the SEP trigger threshold to 15\% in deep space and 18\% on the Martian surface. However, the latter would be associated with an approximately  8-percentage-point higher probability of a false trigger.

\section{Summary and Conclusion}\label{Chapter_Summary}

Sudden radiation exposure from SEP events, both en route to and on the Martian surface, poses a significant health threat to astronauts on future Mars missions. These events not only increase long-term cancer risk but, in extreme cases, can lead to acute radiation syndrome, which can jeopardize the mission. Predicting SEP events is crucial to provide astronauts with sufficient time to move to shelter. However, all available tools are specifically designed for the Earth or Earth-Moon system, and there is currently no consistent way to forecast SEP events for a round-trip mission to Mars.

To address this, we developed a robust and easily implementable nowcasting system for SEP events that is applicable in deep space and on the surface of Mars. Our system is based on dose rate measurements in the MSL/RAD plastic scintillator, which recorded 5 SEP events during the seven-month flight to Mars and 16 SEP events since its landing on Mars on August 6, 2012. Our nowcasting system serves as a reliable last-resort backup when forecasts fail, operating as follows:

The average radiation background is automatically determined from measurements taken over the past five days and is updated daily. Each measurement from the current day is then compared to this radiation background. If a single measurement exceeds the background level by more than 25\%, the system is triggered, and the astronaut is informed of a potential SEP event, prompting them to begin preparing to seek shelter. Based on our nowcasting system using SEPs in deep space, there is an approximately 100\% probability that such a trigger corresponds to an actual SEP event, while on the surface of Mars, this probability is approximately 75\%. A consecutive measurement is then used as a cross-check to verify or falsify the SEP event. If the consecutive measurement also exceeds the SEP trigger threshold, the SEP event is confirmed, and the astronaut is alerted. At this stage, there is near certainty ($\sim$100\%) of the occurrence of an SEP event, both in deep space and on the Martian surface. 

Our study suggests that our nowcasting system can consistently provide astronauts with at least 30 minutes to evade both peak radiation exposure and the majority of the cumulative dose from SEP events. For the largest SEP event measured in deep space on March 7, 2012, an astronaut using our nowcasting system would have had even approximately one and a half days to mitigate $\sim$8.9 mGy, which is almost nine times the annual ICRP dose limit for public exposure. For the largest SEP event ever measured on the surface of Mars on May 20, 2024, an astronaut with our nowcasting system would have had about 50 minutes to mitigate $\sim$1.56 mGy, which is almost 1.6 times the annual ICRP dose limit for public exposure.

\begin{figure}[H]\label{Summary_Plot}
    \centering
    \includegraphics[width=0.9\textwidth]{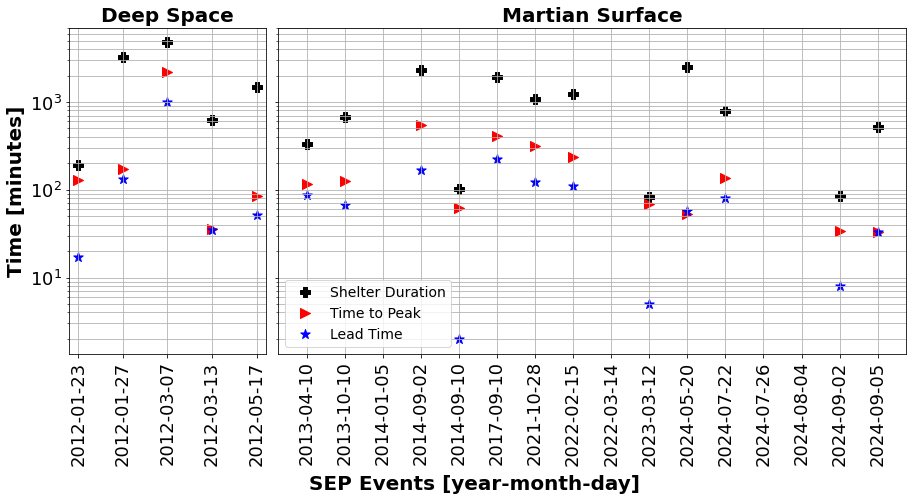}
    \caption{Summary plot of the shelter duration (black), time to peak (red) and lead time (blue) for all SEP events detected by MSL/RAD in deep space (left) and on the Martian surface (right).}
\end{figure}

Additionally, by automatically notifying astronauts when the SEP event has ended, our system also provides information on the required shelter duration, which can range from a few hours to several days, depending on the event. A summary of shelter duration, time to peak, and lead time is depicted in Figure \ref{Summary_Plot}. 

In order to further improve our nowcasting system, it is proposed to lower the trigger threshold for a SEP event from 25\% to as low as 15\% in deep space and 18\% on the Martian surface, even though the latter comes at the cost of a slightly higher FAR. Furthermore, it can be tested whether our available 15-minute data product is the best for nowcasting or if a shorter measurement interval could potentially enable faster nowcasting with still acceptable FAR. The caveat is that the presented results are valid for the statistics of the MSL/RAD plastic scintillator detector. For a future Mars mission, it is recommended that astronauts carry a nowcasting system similar to MSL/RAD. Otherwise, this study should be re-evaluated for the dosimeters they may carry.

\newpage

\section{Appendix: Nowcasting SEP Events with Dose Rate B}\label{SM_NowcastingB}

For the development of our nowcasting system, we utilized the dose rate in plastic scintillator E, which turned out to be the best quantity for nowcasting. In this section, we aim to demonstrate why the dose rate B, measured in the Si detector of MSL/RAD, is suitable for analyzing SEP events but not for nowcasting.

The rover 'Curiosity' is powered by a radioisotope thermoelectric generator (RTG). Dose rate E is not affected by the additional radiation from radioactive decay, as its detection threshold is sufficiently high to exclude the RTG background. In contrast, the measurements of dose rate B include the RTG background. Since future Mars missions are expected to involve astronauts carrying dosimeters without additional radioactive isotopes, we correct dose rate B for the RTG background using the exponential decay law:

\begin{equation}
    B = B_\text{RTG} - B_0\cdot\exp\left(-\frac{\tau}{t}\right) \, .
\end{equation}

\begin{figure}[H]
    \centering
    \includegraphics[width=0.95\textwidth]{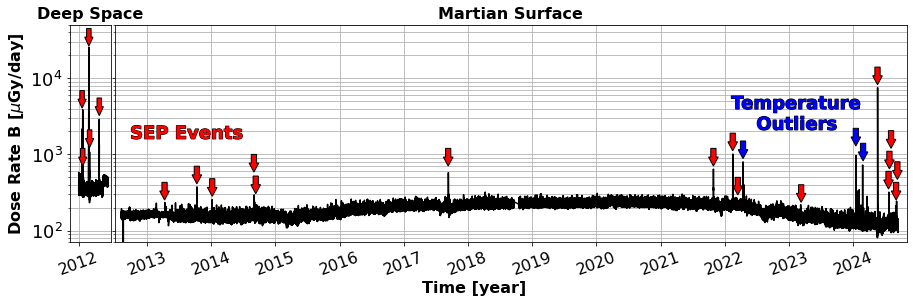}
    \caption{The dose rate B measured by the MSL/RAD in deep space since the launch on November 26, 2011, during the seven-month flight to Mars (left) and on the Martian surface from landing on August 6, 2012, to September 9, 2024 (right). Detected SEP events are indicated by red arrows. Temperature Outliers are denoted with blue arrows.}
    \label{fig_DoseB}
\end{figure}

The measured uncorrected dose rate, including the RTG background, is denoted as RTG as $B_\text{RTG}$. The RTG dose rate at the start of the MSL mission is adjusted from the study by \cite{Zeitlin2016} and is $B_0 = 67\mu$ Gy/Day. The characteristic decay time of Plutonium ($^{238}$Pu) is approximately $\tau = 31175.5$ days. The RTG-corrected dose rate B, measured in the $\sim$15-minute cadences, is shown in Figure \ref{fig_DoseB}. The left panel represents the seven-month flight to Mars from November 26, 2011, to the landing on the Martian surface on August 6, 2012. The right panel shows the measurements on the surface of Mars up to September 9, 2024.

\begin{figure}[H]
    \centering
    \includegraphics[width=0.95\textwidth]{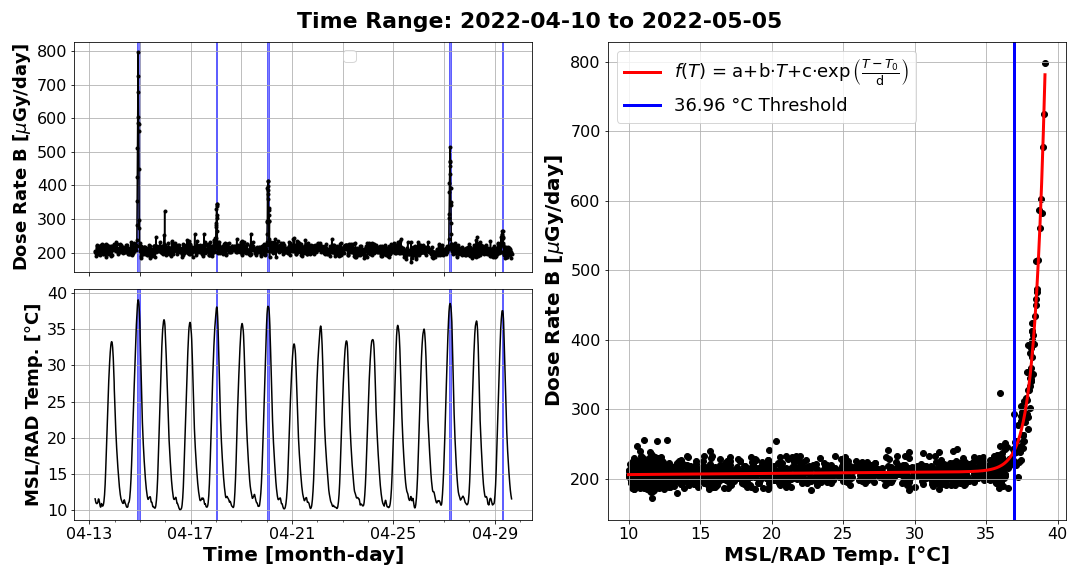}
    \caption{Analysis of the temperature impact on dose rate B for the period from April 10, 2022, to May 5, 2022. On the left, dose rate B and the internal MSL/RAD temperature are plotted versus time. Temperature ranges exceeding 36.96°C are marked in blue and precisely overlap with the spikes in dose rate B. On the right, the fit with a model function consisting of a linear and an exponential contribution is shown to determine the temperature threshold at which temperature effects occur, resulting in 36.96°C.}
    \label{fig_TempStudy}
\end{figure}

Unlike dose rate E, dose rate B exhibits spikes unrelated to SEP events. In Figure \ref{fig_DoseB}, the largest spikes are marked with blue arrows and appear only after 2022, particularly during the Martian summer from April 10, 2022, to May 5, 2022, and from January 17, 2024, to February 7, 2024. These spikes can be attributed to the increased internal temperature of MSL/RAD, which, while expected after more than 13 years of operation, must still be considered in our analysis. Figure \ref{fig_TempStudy} (left) shows dose rate B during the period from April 10 to May 5, 2022, linked with the internal MSL/RAD temperature. The spikes occur when the internal temperature is slightly higher than usual. We flag the time ranges affected by temperature by calculating a temperature threshold at which these effects emerge. For this, we plot dose rate B against the internal MSL/RAD temperature (see Figure \ref{fig_TempStudy}, right) and fit a model function to the data, consisting of both linear and exponential contributions:

\begin{equation}
    f(T) = \text{a}+\text{b}\cdot T+\text{c}\cdot\exp\left(\frac{T-T_0}{\text{d}}\right) \, .
\end{equation}

The fit is shown as a red curve. The corresponding fit parameters are a=204.55 $\mu$Gy/day, b=0.17 $\mu$Gy/day/°C, c=$4.23\cdot 10^{-15} \mu$Gy/day, T$_0$= 10.2 °C and d=0.73 °C. We define the impact of temperature effects when dose rate B exceeds the linear contribution of the fitted function by more than 5\%. The increase of more than 5\% is marked by a blue vertical line in Figure \ref{fig_TempStudy} and is determined to occur at 36.96 °C. During the second time period, from January 17, 2024, to February 7, 2024, where temperature-related spikes in dose rate B are particularly pronounced, the same analysis establishes the threshold at 35.77 °C. For nowcasting, we mask these temperature-related spikes. However, for real-life nowcasting, it is important to keep in mind that such effects may occur during a prolonged Mars mission.

\begin{figure}[H]
    \centering
    \includegraphics[width=0.9\textwidth]{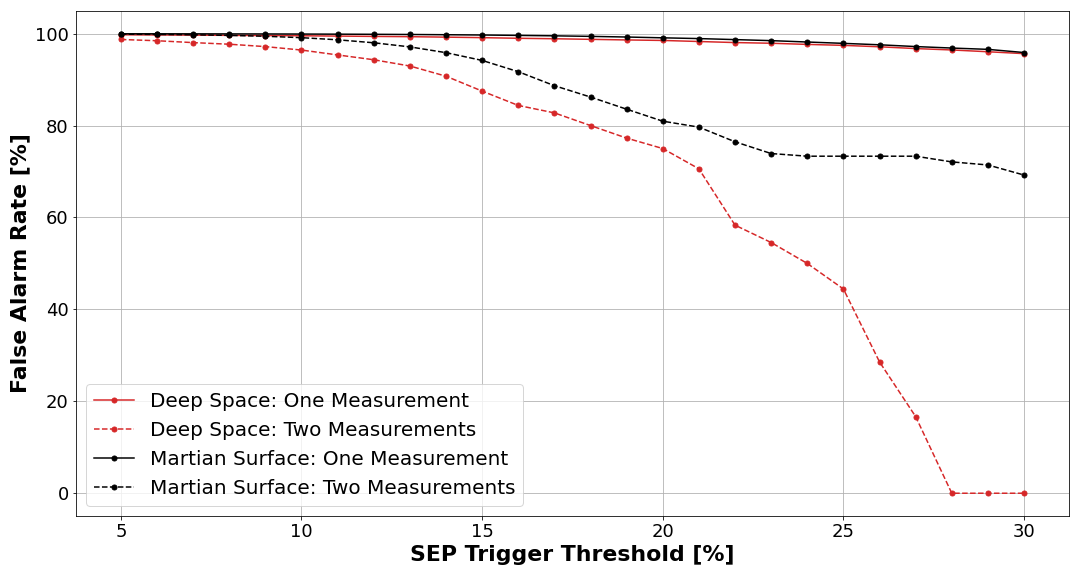}
    \caption{Similar to the study and representation of the FAR in the main paper, but using dose rate B. The error bars are omitted for clarity.}
    \label{fig_FalseAlarm_B}
\end{figure}

Using the same methodology introduced in the main paper, we determine the FAR with the dose rate B, corrected for the RTG influence and temperature outliers. The results are plotted in Figure \ref{fig_FalseAlarm_B} without error bars for clarity. 

 Nowcasting based on single measurements of the dose rate B has a FAR $\gtrsim$95\% in both deep space and on the Martian surface, even with a trigger threshold set 30\% above the background. Nowcasting in deep space based on two measurements drops to around 0\% at a trigger threshold of 28\%, but on the surface of Mars, it still remains $\sim$70\%, even with a trigger threshold of 30\%. As a result, nowcasting with dose rate B is not reliable. The reason why nowcasting is possible with dose rate E but not with dose rate B is that the geometric factor of the E detector is significantly larger than that of the B detector. Consequently, the statistics are significantly better, and the measurements are less susceptible to fluctuations. The larger fluctuations in dose rate B compared to dose rate E are particularly evident in Figure \ref{fig_DoseB}. Furthermore, the dose rate B is generally lower than dose rate E. The trigger thresholds for an SEP event, defined relative to the background, are therefore smaller for dose rate B compared to dose rate E, leading to a more sensitive triggering of the nowcasting system.
 
Although nowcasting with dose rate B is not effective, SEP events can still be analyzed using dose rate B.  The results for shelter duration, peak time, lead time, and total SEP dose using dose rate B are provided in Table \ref{tab:Results_B}.

\begin{table}[H]
    \centering
    \renewcommand{\arraystretch}{1.5} 
    \caption{Analysis of shelter duration, time to peak, lead time, and total SEP dose for the SEP events measured in the dose rate B with MSL/RAD. The events are categorized into those measured in deep space and on the surface of Mars. The grey-highlighted rows correspond to the largest SEP event measured in deep space and on the surface of Mars, respectively.}
    \begin{tabular}{@{\extracolsep{0pt}}
    >{\centering\arraybackslash}p{0.1cm}>{\centering\arraybackslash}p{2.78cm} | >{\centering\arraybackslash}p{3.03cm} | >{\centering\arraybackslash}p{3.03cm} >{\centering\arraybackslash}p{3.03cm} | >{\centering\arraybackslash}p{2.32cm}
    }
    \toprule
     & \textbf{SEP Event [year-month-day]} & \textbf{Shelter Duration [day\space\space hour:minute]} & \textbf{Time to Peak \space\space\space\space[day\space\space  hour:minute]} & \textbf{Lead Time\space\space\space\space\space\space\space\space [day\space\space hour:minute]} & \textbf{Total SEP Dose [$\mu$Gy]}\\
    \midrule
    \parbox[t]{1mm}{\multirow{5}{*}{\rotatebox[origin=c]{90}{\textbf{DEEP SPACE}}}} 
    & 2012-01-23 & 02d\space\space18h:37m & 00d\space\space11h:14m & 00d\space\space05h:42m & 1116.27  \\
    & 2012-01-27 & 01d\space\space23h:57m & 00d\space\space04h:10m & 00d\space\space02h:37m & 1697.14 \\
    & \cellcolor{gray!20}2012-03-07 & \cellcolor{gray!20}04d\space\space09h:43m & \cellcolor{gray!20}01d\space\space13h:56m & \cellcolor{gray!20}00d\space\space20h:58m & \cellcolor{gray!20}14260.81 \\
    & 2012-03-13 & 00d\space\space16h:24m & 00d\space\space00h:32m & 00d\space\space00h:02m & 82.83   \\
    & 2012-05-17 & 01d\space\space03h:26m & 00d\space\space01h:42m & 00d\space\space01h:08m & 813.42 \\
    \midrule
    \parbox[t]{1mm}{\multirow{16}{*}{\rotatebox[origin=c]{90}{\textbf{MARTIAN SURFACE}}}} 
    & 2013-04-10 & 00d\space\space06h:07m & 00d\space\space02h:02m & 00d\space\space00h:15m & 16.43   \\ 
    & 2013-10-10 & 00d\space\space15h:28m & 00d\space\space01h:58m & 00d\space\space01h:07m & 103.07   \\
    & 2014-01-05 &           &           &           &         \\
    & 2014-09-02 & 02d\space\space21h:32m & 00d\space\space04h:42m & 00d\space\space04h:42m & 137.69  \\
    & 2014-09-10 &          &   &   &    \\
    & 2017-09-10 & 02d\space\space04h:30m & 00d\space\space06h:11m & 00d\space\space04h:55m & 347.55  \\
    & 2021-10-28 & 00d\space\space23h:37m & 00d\space\space05h:03m & 00d\space\space02h:23m & 190.67  \\
    & 2022-02-15 & 01d\space\space16h:17m & 00d\space\space03h:46m & 00d\space\space02h:15m & 370.78  \\
    & 2022-03-14 &           &           &           &         \\
    & 2023-03-12 &  &  &  &    \\
    & \cellcolor{gray!20}2024-05-20 & \cellcolor{gray!20}02d\space\space03h:48m & \cellcolor{gray!20}00d\space\space00h:53m & \cellcolor{gray!20}00d\space\space01h:02m & \cellcolor{gray!20}1639.83 \\
    & 2024-07-22 & 01d\space\space08h:23m & 00d\space\space02h:16m & 00d\space\space01h:42m & 75.65   \\
    & 2024-07-26 &           &           &           &         \\
    & 2024-08-04 &           &           &           &         \\
    & 2024-09-02 & 00d\space\space08h:34m & 00d\space\space00h:51m & 00d\space\space00h:44m & 11.23    \\
    & 2024-09-05 & 00d\space\space20h:25m & 00d\space\space00h:50m & 00d\space\space00h:55m & 36.18   \\
    \bottomrule
    \end{tabular}
    \label{tab:Results_B}
\end{table}

\section{Appendix: Background Fits for SEP Events Measured by MSL/RAD}\label{SM_Background}

For the development of our nowcasting system, we determined the onset of SEP events measured by MSL/RAD with respect to the dose rate background through linear fits over a five-day period prior to each SEP event. Table \ref{tab_BG_Dose} contains our determined background values, along with the standard deviation, for dose rate B and dose rate E. The data is categorized by SEP events measured in deep space and on the surface of Mars. The grey-highlighted rows correspond to the largest measured event in deep space and on the surface of Mars. The timestamps in the first column refer to the SEP events, while the background dose rate fits in columns two and three correspond to the period preceding the respective timestamp.

\begin{table}[H]
    \centering
    \renewcommand{\arraystretch}{1.5} 
    \caption{Background dose rates in detector B and detector E fitted in five-day intervals prior to the SEP events in deep space and on the Martian surface. The grey-highlighted rows correspond to the largest SEP event measured in deep space and on the surface of Mars, respectively}
    \begin{tabular}{@{\extracolsep{0pt}}
    >{\centering\arraybackslash}p{0.1cm}
    >{\centering\arraybackslash}p{4cm} | 
    >{\centering\arraybackslash}p{4.5cm} 
    >{\centering\arraybackslash}p{4.5cm} }
    \toprule
 
     & \textbf{SEP Event [year-month-day]} & \textbf{Background Dose Rate B [$\mu$Gy/day]} & \textbf{Background Dose Rate E [$\mu$Gy/day]} \\
    \midrule
    \parbox[t]{1mm}{\multirow{5}{*}{\rotatebox[origin=c]{90}{\textbf{DEEP SPACE}}}} 
    & 2012-01-23 & 369.04 $\pm$ 1.41 & 435.07 $\pm$ 0.43 \\
    & 2012-01-27 & 357.53 $\pm$ 3.42 & 389.79 $\pm$ 0.93 \\
    & \cellcolor{gray!20}2012-03-07 & \cellcolor{gray!20}341.48 $\pm$ 1.14 & \cellcolor{gray!20}450.42 $\pm$ 0.76 \\
    & 2012-03-13 & 314.75 $\pm$ 4.37 & 370.08 $\pm$ 1.11 \\
    & 2012-05-17 & 348.80 $\pm$ 1.17 & 457.08 $\pm$ 0.48 \\
    \midrule
    \parbox[t]{1mm}{\multirow{16}{*}{\rotatebox[origin=c]{90}{\textbf{MARTIAN SURFACE}}}} & 2013-04-10 & 161.02 $\pm$ 0.33 &  210.01 $\pm$ 0.19 \\
    & 2013-10-10 & 154.26 $\pm$ 0.28 & 209.99 $\pm$ 0.11 \\
    & 2014-01-05 & 149.16 $\pm$ 0.26 & 204.31 $\pm$ 0.11 \\
    & 2014-09-02 & 158.08 $\pm$ 0.29 & 215.26 $\pm$ 0.13 \\
    & 2014-09-10 & 144.80 $\pm$ 0.94 & 196.84 $\pm$ 0.27 \\
    & 2017-09-10 & 200.66 $\pm$ 0.34 & 269.28 $\pm$ 0.23 \\
    & 2021-10-28 & 214.58 $\pm$ 0.33 & 287.69 $\pm$ 0.18 \\
    & 2022-02-15 & 208.89 $\pm$ 0.33 & 279.11 $\pm$ 0.19 \\
    & 2022-03-14 & 217.94 $\pm$ 0.62 & 291.12 $\pm$ 0.29 \\
    & 2023-03-12 & 144.50 $\pm$ 0.37 & 194.95 $\pm$ 0.16 \\
    & \cellcolor{gray!20}2024-05-20 & \cellcolor{gray!20}121.81 $\pm$ 0.24 & \cellcolor{gray!20}164.35 $\pm$ 0.09 \\
    & 2024-07-22 & 118.91 $\pm$ 0.28 & 163.08 $\pm$ 0.12 \\
    & 2024-07-26 & 122.62 $\pm$ 1.00 & 164.91 $\pm$ 0.31 \\
    & 2024-08-04 & 120.07 $\pm$ 0.38 & 165.29 $\pm$ 0.19 \\
    & 2024-09-02 & 121.45 $\pm$ 0.31 & 164.82 $\pm$ 0.15 \\
    & 2024-10-05 & 127.46 $\pm$ 1.65 & 169.88 $\pm$ 0.55 \\
    \bottomrule
    \end{tabular}
    \label{tab_BG_Dose}
\end{table}

\section{Appendix: Threshold Study with Dose Rate E} \label{SM_ThresholdStudy}

\begin{figure}[H]
    \centering
    \includegraphics[width=1.0\textwidth]{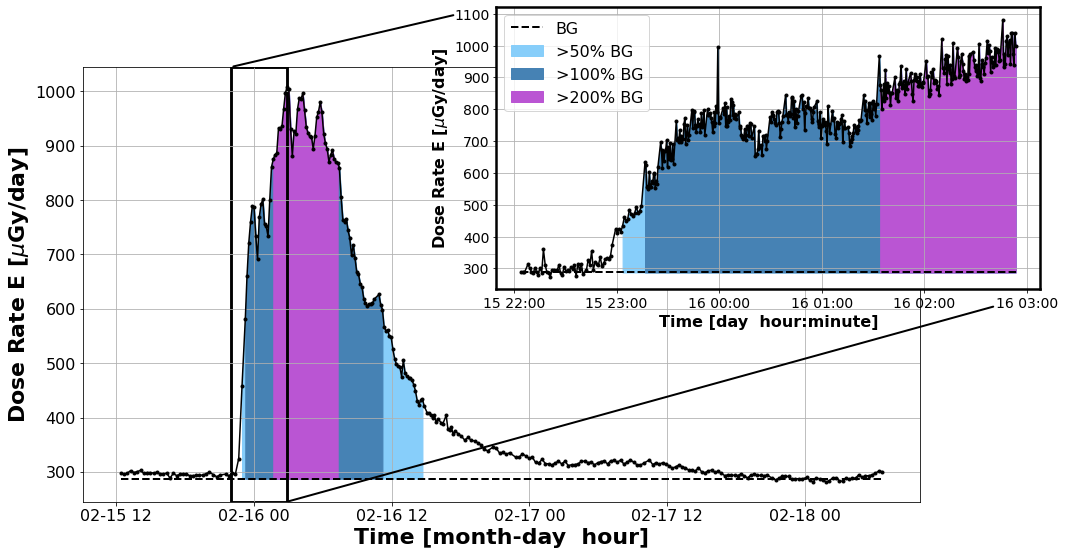}
    \caption{Dose rate E plotted versus time. The main plot shows the entire SEP event on February 25, 2022, while the subplot provides a zoomed-in view of the high-resolution dose rate E from the onset to the peak of the event. The black dashed line represents the fitted background (BG). In the subplot, the times required to reach the thresholds of 50\%, 100\%, and 200\% above background are highlighted with different colors. Similarly, in the main plot, the time intervals during which the dose rates remain above these thresholds are marked using the same color scheme.}
    \label{fig_Threshold}
\end{figure}

In the main paper, we defined the shelter duration as the period during which dose rate E remains more than 25\% above the background level. Here, we also examine the time intervals during which the dose rate E exceeds 50\%, 100\%, and 200\% above the background level, respectively. These durations represent potential shelter times for an astronaut who does not seek immediate protection at the onset of the SEP event or who exits the shelter before the dose rate E returns to background levels.The main plot in Figure \ref{fig_Threshold} highlights the time intervals exceeding each respective background threshold. The black dotted line indicates the fitted background.

Using the high-resolution dose rate E, we further analyze the time required from the onset of the event to reach the 50\%, 100\%, and 200\% thresholds above the background. The subplot of Figure \ref{fig_Threshold} offers a zoomed-in view of the high-resolution dose rate E, from the onset to the peak of the SEP event. The moments when the dose rate exceeds the respective background thresholds are marked in the same color scheme as the main plot. The majority of the dose rate E must exceed the respective thresholds in order to be considered as a trigger for the threshold. A single measurement does not qualify as a threshold trigger. An example of such a case can be seen in the subplot of Figure \ref{fig_Threshold}, where just a measurement at 2022-02-16 00:00 exceeds the 200\% threshold. This outlier is excluded to prevent triggering the thresholds due to e.g. heavy ions unrelated to SEP events. This analysis aims to determine whether an astronaut could potentially avoid not only the peak dose rate of an SEP event (referred to as 'time to peak' in the main paper) but also have sufficient time to seek shelter at an earlier stage.

In the main paper, we defined the lead time as the duration required for an astronaut to avoid at least 90\% of the total SEP dose. Here, we further analyze the times available for avoiding at least 80\% and 50\% of the total SEP dose. For cases where insufficient lead time exists to avoid 90\% of the dose, we assess whether it is still possible to mitigate at least 80\% or, at minimum, 50\% of the total dose. The results of all threshold analyses are summarized in Table \ref{tab_threshold}. No entries in the 50\%, 100\%, and 200\% above background columns indicate that the dose rate E does not exceed these thresholds. No entries in the last two columns indicate that the SEP event is so minor that it does not even exceed the onset threshold of 25\% above background.

\begin{sidewaystable}
    \centering
    \renewcommand{\arraystretch}{1.5} 
    \caption{Threshold study for SEP events in deep space and on the surface of Mars. The analysis records the durations during which the dose rate exceeds 50\%, 100\%, and 200\% of the background level. Additionally, the times from onset to the first exceedance of each respective threshold are documented. The last two columns specify the available times for an astronaut to avoid at least 80\% and 50\% of the total SEP dose.}
    \begin{tabular}{@{\extracolsep{0pt}}
    >{\centering\arraybackslash}p{0cm}>{\centering\arraybackslash}p{2.8cm} | 
    >{\centering\arraybackslash}p{2.1cm} >{\centering\arraybackslash}p{2.1cm} >{\centering\arraybackslash}p{2.1cm} |
    >{\centering\arraybackslash}p{2.1cm} >{\centering\arraybackslash}p{2.1cm} >{\centering\arraybackslash}p{2.1cm} |
    >{\centering\arraybackslash}p{2.1cm} >{\centering\arraybackslash}p{2.1cm} 
    }
    \toprule
    & &  \multicolumn{3}{c|}{\textbf{Duration [day\space\space hour:minute] for dose rate E to be...}} & \multicolumn{3}{c|}{\textbf{Time [day\space\space hour:minute] for dose rate E to reach...}} & \multicolumn{2}{c}{\textbf{Time[day\space\space hour:minute] to avoid}} \\

     & \textbf{SEP Event [year-month-day]} & \textbf{$>$ 50\% above BG} & \textbf{$>$ 100\% above BG} & \textbf{$>$ 200\% above BG} & \textbf{$>$ 50\% above BG} &  \textbf{$>$ 100\% above BG} & \textbf{$>$ 200\% above BG} & \textbf{80\% of total SEP dose} & \textbf{50\% of total SEP dose} \\
    \midrule
    \parbox[t]{1mm}{\multirow{5}{*}{\rotatebox[origin=c]{90}{\textbf{DEEP SPACE}}}} 
    & 2012-01-23 &                      &                      &                      &                      &                      &                      & 00\space\space 00:35 & 00\space\space 01:33 \\
    & 2012-01-27 & 01\space\space 06:07 & 00\space\space 20:06 & 00\space\space 13:20 & 00\space\space 00:05 & 00\space\space 00:17 & 00\space\space 00:35 & 00\space\space 03:23 & 00\space\space 07:49 \\
    & 2012-03-07 & 03\space\space 01:15 & 01\space\space 16:49 & 00\space\space 02:08 & 00\space\space 01:34 & 00\space\space 02:13 & 00\space\space 06:06 & 00\space\space 23:45 & 01\space\space 11:32 \\
    & 2012-03-13 & 00\space\space 04:52 &                      &                      & 00\space\space 00:03 &                      &                      & 00\space\space 01:11 & 00\space\space 03:16 \\
    & 2012-05-17 & 00\space\space 17:09 & 00\space\space 13:22 & 00\space\space 09:59 & 00\space\space 00:17 & 00\space\space 00:17 & 00\space\space 00:17 & 00\space\space 01:30 & 00\space\space 04:32 \\
    \midrule
    \parbox[t]{1mm}{\multirow{16}{*}{\rotatebox[origin=c]{90}{\textbf{MARTIAN SURFACE}}}} 
    & 2013-04-10 &                      &                      &                      &                      &                      &                      & 00\space\space 01:28 & 00\space\space 02:13 \\ 
    & 2013-10-10 & 00\space\space 05:56 &                      &                      & 00\space\space 00:08 &                      &                      & 00\space\space 01:58 & 00\space\space 03:41 \\
    & 2014-01-05 &                      &                      &                      &                      &                      &                      &                      &                      \\
    & 2014-08-31 & 00\space\space 07:55 &                      &                      & 00\space\space 03:47 &                      &                      & 00\space\space 05:46 & 00\space\space 14:21 \\
    & 2014-09-10 &                      &                      &                      &                      &                      &                      & 00\space\space 00:15 & 00\space\space 01:28 \\
    & 2017-09-10 & 00\space\space 22:05 & 00\space\space 10:06 &                      & 00\space\space 04:36 & 00\space\space 05:31 &                      & 00\space\space 05:54 & 00\space\space 12:09 \\
    & 2021-10-28 & 00\space\space 11:21 & 00\space\space 05:54 &                      & 00\space\space 00:25 & 00\space\space 01:46 &                      & 00\space\space 03:14 & 00\space\space 06:29 \\
    & 2022-02-15 & 00\space\space 15:50 & 00\space\space 12:23 & 00\space\space 05:53 & 00\space\space 00:06 & 00\space\space 00:19 & 00\space\space 02:36 & 00\space\space 03:15 & 00\space\space 06:42 \\
    & 2022-03-14 &                      &                      &                      &                      &                      &                      &                      &                      \\
    & 2023-03-12 &                      &                      &                      &                      &                      &                      & 00\space\space 00:13 & 00\space\space 01:11 \\
    & 2023-05-19 & 01\space\space 11:04 & 01\space\space 04:48 & 00\space\space 20:16 & 00\space\space 00:02 & 00\space\space 00:04 & 00\space\space 00:08 & 00\space\space 01:29 & 00\space\space 04:24 \\
    & 2024-07-22 & 00\space\space 08:08 & 00\space\space 01:25 &                      & 00\space\space 00:19 & 00\space\space 01:25 &                      & 00\space\space 02:11 & 00\space\space 04:41 \\
    & 2024-07-26 &                      &                      &                      &                      &                      &                      &                      &                      \\
    & 2024-08-04 &                      &                      &                      &                      &                      &                      &                      &                      \\
    & 2024-09-02 &                      &                      &                      &                      &                      &                      & 00\space\space 00:17 & 00\space\space 00:41 \\
    & 2024-10-09 & 00\space\space 04:49 &                      &                      & 00\space\space 00:20 &                      &                      & 00\space\space 01:09 & 00\space\space 04:20 \\
    \bottomrule
    \end{tabular}
    \label{tab_threshold}
\end{sidewaystable}

\newpage

\section*{Open Research}
The data used in this study are archived in the NASA Planetary Data System's Planetary Plasma Interactions Node at the University of California, Los Angeles. The archival volume includes the full binary raw data files, detailed descriptions of the structures, and higher-level data products in human-readable form. The binary RAD EDR data are archived under \cite{Peterson2013}, while the human-readable RAD RDR data are archived under \cite{Rafkin2013}. The data used to produce the figures in the study, as well as the temperature impact analysis and threshold evaluation in the supplementary material, are available as open access in Zenodo \cite{loewe_2025_14801849}. More information about the structure of the files is contained within the files themselves.

\section*{Acknowledgments}
RAD is supported by NASA (HEOMD) under Jet Propulsion Laboratory (JPL) subcontract 1273039 to Southwest Research Institute and in Germany by the German Aerospace Center (DLR) and DLR's Space Administration grant 50QM1701 to the Christian Albrechts-Universität zu Kiel. 
JG thanks the support from National Natural Science Foundation of China (Grant Nos. 42188101, 42130204, 42474221).

%\acknowledgments
%Enter acknowledgments, including your data availability statement, here.

%% ------------------------------------------------------------------------ %%
%% References and Citations

%%%%%%%%%%%%%%%%%%%%%%%%%%%%%%%%%%%%%%%%%%%%%%%
%
% \bibliography{<name of your .bib file>} don't specify the file extension
%
% don't specify bibliographystyle
%%%%%%%%%%%%%%%%%%%%%%%%%%%%%%%%%%%%%%%%%%%%%%%
%\printbibliography

\bibliography{sample.bib}

\begin{thebibliography}{}

\bibitem [\protect \citeauthoryear {%
Appel%
\ \protect \BOthers {.}}{%
Appel%
\ \protect \BOthers {.}}{%
{\protect \APACyear {2018}}%
}]{%
appel2018detecting}
\APACinsertmetastar {%
appel2018detecting}%
\begin{APACrefauthors}%
Appel, J.%
, K{\"o}ehler, J.%
, Guo, J.%
, Ehresmann, B.%
, Zeitlin, C.%
, Matthi{\"a}, D.%
\BDBL {}others%
\end{APACrefauthors}%
\unskip\
\newblock
\APACrefYearMonthDay{2018}{}{}.
\newblock
{\BBOQ}\APACrefatitle {Detecting upward directed charged particle fluxes in the
  mars science laboratory radiation assessment detector} {Detecting upward
  directed charged particle fluxes in the mars science laboratory radiation
  assessment detector}.{\BBCQ}
\newblock
\APACjournalVolNumPages{Earth and Space Science}{5}{1}{2--18}.
\newblock
\begin{APACrefDOI} \doi{https://doi.org/10.1002/2016EA000240} \end{APACrefDOI}
\PrintBackRefs{\CurrentBib}

\bibitem [\protect \citeauthoryear {%
Badhwar%
\ \BBA {} O'Neill%
}{%
Badhwar%
\ \BBA {} O'Neill%
}{%
{\protect \APACyear {1994}}%
}]{%
Badhwar1994}
\APACinsertmetastar {%
Badhwar1994}%
\begin{APACrefauthors}%
Badhwar, G\BPBI D.%
\BCBT {}\ \BBA {} O'Neill, P\BPBI M.%
\end{APACrefauthors}%
\unskip\
\newblock
\APACrefYearMonthDay{1994}{{\APACmonth{10}}}{}.
\newblock
{\BBOQ}\APACrefatitle {Long-term modulation of Galactic Cosmic Radiation and
  its model for space exploration} {Long-term modulation of galactic cosmic
  radiation and its model for space exploration}.{\BBCQ}
\newblock
\APACjournalVolNumPages{Advances in Space Research}{14}{10}{749--757}.
\newblock
\begin{APACrefDOI} \doi{10.1016/0273-1177(94)90537-1} \end{APACrefDOI}
\PrintBackRefs{\CurrentBib}

\bibitem [\protect \citeauthoryear {%
{Barcellos-Hoff}%
\ \protect \BOthers {.}}{%
{Barcellos-Hoff}%
\ \protect \BOthers {.}}{%
{\protect \APACyear {2015}}%
}]{%
Barcellos-Hoff2015}
\APACinsertmetastar {%
Barcellos-Hoff2015}%
\begin{APACrefauthors}%
{Barcellos-Hoff}, M\BPBI H.%
, {Blakely}, E\BPBI A.%
, {Burma}, S\BPBI e.%
, {Fornace}, A\BPBI J.%
, {Gerson}, S.%
, {Hlatky}, L.%
\BDBL {}{Weil}, M\BPBI M.%
\end{APACrefauthors}%
\unskip\
\newblock
\APACrefYearMonthDay{2015}{Jul}{}.
\newblock
{\BBOQ}\APACrefatitle {{Concepts and challenges in cancer risk prediction for
  the space radiation environment}} {{Concepts and challenges in cancer risk
  prediction for the space radiation environment}}.{\BBCQ}
\newblock
\APACjournalVolNumPages{Life Sciences and Space Research}{6}{}{92-103}.
\newblock
\begin{APACrefDOI} \doi{10.1016/j.lssr.2015.07.006} \end{APACrefDOI}
\PrintBackRefs{\CurrentBib}

\bibitem [\protect \citeauthoryear {%
Baselet%
, Rombouts%
, Benotmane%
, Baatout%
\BCBL {}\ \BBA {} Aerts%
}{%
Baselet%
\ \protect \BOthers {.}}{%
{\protect \APACyear {2016}}%
}]{%
Baselet2016}
\APACinsertmetastar {%
Baselet2016}%
\begin{APACrefauthors}%
Baselet, B.%
, Rombouts, C.%
, Benotmane, A\BPBI M.%
, Baatout, S.%
\BCBL {}\ \BBA {} Aerts, A.%
\end{APACrefauthors}%
\unskip\
\newblock
\APACrefYearMonthDay{2016}{}{}.
\newblock
{\BBOQ}\APACrefatitle {Cardiovascular diseases related to ionizing radiation:
  The risk of low-dose exposure (Review)} {Cardiovascular diseases related to
  ionizing radiation: The risk of low-dose exposure (review)}.{\BBCQ}
\newblock
\APACjournalVolNumPages{International Journal of Molecular
  Medicine}{38}{}{1623--1641}.
\newblock
\begin{APACrefDOI} \doi{10.3892/ijmm.2016.2777} \end{APACrefDOI}
\PrintBackRefs{\CurrentBib}

\bibitem [\protect \citeauthoryear {%
Boerma%
\ \protect \BOthers {.}}{%
Boerma%
\ \protect \BOthers {.}}{%
{\protect \APACyear {2015}}%
}]{%
Boerma2015}
\APACinsertmetastar {%
Boerma2015}%
\begin{APACrefauthors}%
Boerma, M.%
, Nelson, G\BPBI A.%
, Sridharan, V.%
, Mao, X\BPBI W.%
, Koturbash, I.%
\BCBL {}\ \BBA {} Hauer-Jensen, M.%
\end{APACrefauthors}%
\unskip\
\newblock
\APACrefYearMonthDay{2015}{}{}.
\newblock
{\BBOQ}\APACrefatitle {Space radiation and cardiovascular disease risk} {Space
  radiation and cardiovascular disease risk}.{\BBCQ}
\newblock
\APACjournalVolNumPages{World Journal of Cardiology}{7}{12}{882--888}.
\newblock
\begin{APACrefDOI} \doi{10.4330/wjc.v7.i12.882} \end{APACrefDOI}
\PrintBackRefs{\CurrentBib}

\bibitem [\protect \citeauthoryear {%
Crosby%
\ \protect \BOthers {.}}{%
Crosby%
\ \protect \BOthers {.}}{%
{\protect \APACyear {2012}}%
}]{%
COMESEP}
\APACinsertmetastar {%
COMESEP}%
\begin{APACrefauthors}%
Crosby, N.%
, Veronig, A.%
, Robbrecht, E.%
, Vrsnak, B.%
, Vennerstrom, S.%
, Malandraki, O.%
\BDBL {}Consortium, C.%
\end{APACrefauthors}%
\unskip\
\newblock
\APACrefYearMonthDay{2012}{11}{}.
\newblock
{\BBOQ}\APACrefatitle {Forecasting the space weather impact: The COMESEP
  project} {Forecasting the space weather impact: The comesep project}.{\BBCQ}
\newblock
\APACjournalVolNumPages{AIP Conference Proceedings}{1500}{}{159-164}.
\newblock
\begin{APACrefDOI} \doi{10.1063/1.4768760} \end{APACrefDOI}
\PrintBackRefs{\CurrentBib}

\bibitem [\protect \citeauthoryear {%
Cucinotta%
, Wang%
\BCBL {}\ \BBA {} Huff%
}{%
Cucinotta%
\ \protect \BOthers {.}}{%
{\protect \APACyear {2009}}%
}]{%
inbook}
\APACinsertmetastar {%
inbook}%
\begin{APACrefauthors}%
Cucinotta, F.%
, Wang, H.%
\BCBL {}\ \BBA {} Huff, J.%
\end{APACrefauthors}%
\unskip\
\newblock
\APACrefYearMonthDay{2009}{05}{}.
\newblock
{\BBOQ}\APACrefatitle {Risk of Acute or Late Central Nervous System Effects
  from Radiation Exposure} {Risk of acute or late central nervous system
  effects from radiation exposure}.{\BBCQ}
\newblock
\BIn{} (\BPG~191-212).
\PrintBackRefs{\CurrentBib}

\bibitem [\protect \citeauthoryear {%
{Cucinotta}%
, {Alp}%
, {Sulzman}%
\BCBL {}\ \BBA {} {Wang}%
}{%
{Cucinotta}%
\ \protect \BOthers {.}}{%
{\protect \APACyear {2014}}%
}]{%
Cucinotta2014}
\APACinsertmetastar {%
Cucinotta2014}%
\begin{APACrefauthors}%
{Cucinotta}, F\BPBI A.%
, {Alp}, M.%
, {Sulzman}, F\BPBI M.%
\BCBL {}\ \BBA {} {Wang}, M.%
\end{APACrefauthors}%
\unskip\
\newblock
\APACrefYearMonthDay{2014}{Jul}{}.
\newblock
{\BBOQ}\APACrefatitle {{Space radiation risks to the central nervous system}}
  {{Space radiation risks to the central nervous system}}.{\BBCQ}
\newblock
\APACjournalVolNumPages{Life Sciences and Space Research}{2}{}{54-69}.
\newblock
\begin{APACrefDOI} \doi{10.1016/j.lssr.2014.06.003} \end{APACrefDOI}
\PrintBackRefs{\CurrentBib}

\bibitem [\protect \citeauthoryear {%
{Cucinotta}%
\ \BBA {} {Durante}%
}{%
{Cucinotta}%
\ \BBA {} {Durante}%
}{%
{\protect \APACyear {2006}}%
}]{%
Cucinotta2006}
\APACinsertmetastar {%
Cucinotta2006}%
\begin{APACrefauthors}%
{Cucinotta}, F\BPBI A.%
\BCBT {}\ \BBA {} {Durante}, M.%
\end{APACrefauthors}%
\unskip\
\newblock
\APACrefYearMonthDay{2006}{May}{}.
\newblock
{\BBOQ}\APACrefatitle {{Cancer risk from exposure to galactic cosmic rays:
  implications for space exploration by human beings.}} {{Cancer risk from
  exposure to galactic cosmic rays: implications for space exploration by human
  beings.}}{\BBCQ}
\newblock
\APACjournalVolNumPages{Lancet Oncol.}{7}{431-436}{682-688}.
\newblock
\begin{APACrefDOI} \doi{10.1016/S1470-2045(06)70695-7} \end{APACrefDOI}
\PrintBackRefs{\CurrentBib}

\bibitem [\protect \citeauthoryear {%
Cucinotta%
, Wu%
, Shavers%
\BCBL {}\ \BBA {} George%
}{%
Cucinotta%
\ \protect \BOthers {.}}{%
{\protect \APACyear {2003}}%
}]{%
cucinotta2003radiation}
\APACinsertmetastar {%
cucinotta2003radiation}%
\begin{APACrefauthors}%
Cucinotta, F\BPBI A.%
, Wu, H.%
, Shavers, M\BPBI R.%
\BCBL {}\ \BBA {} George, K.%
\end{APACrefauthors}%
\unskip\
\newblock
\APACrefYearMonthDay{2003}{}{}.
\newblock
{\BBOQ}\APACrefatitle {Radiation dosimetry and biophysical models of space
  radiation effects} {Radiation dosimetry and biophysical models of space
  radiation effects}.{\BBCQ}
\newblock
\APACjournalVolNumPages{Gravitational and Space Biology}{16}{2}{11--19}.
\newblock
\begin{APACrefURL}
  \url{https://www.researchgate.net/publication/10580019_Radiation_dosimetry_and_biophysical_models_of_space_radiation_effects}
  \end{APACrefURL}
\PrintBackRefs{\CurrentBib}

\bibitem [\protect \citeauthoryear {%
Desai%
\ \BBA {} Giacalone%
}{%
Desai%
\ \BBA {} Giacalone%
}{%
{\protect \APACyear {2016}}%
}]{%
Desai2016}
\APACinsertmetastar {%
Desai2016}%
\begin{APACrefauthors}%
Desai, M.%
\BCBT {}\ \BBA {} Giacalone, J.%
\end{APACrefauthors}%
\unskip\
\newblock
\APACrefYearMonthDay{2016}{09}{}.
\newblock
{\BBOQ}\APACrefatitle {Large gradual solar energetic particle events} {Large
  gradual solar energetic particle events}.{\BBCQ}
\newblock
\APACjournalVolNumPages{Living Reviews in Solar Physics}{13}{1}{3}.
\newblock
\begin{APACrefURL} \url{https://doi.org/10.1007/s41116-016-0002-5}
  \end{APACrefURL}
\newblock
\begin{APACrefDOI} \doi{10.1007/s41116-016-0002-5} \end{APACrefDOI}
\PrintBackRefs{\CurrentBib}

\bibitem [\protect \citeauthoryear {%
Diaz%
\ \protect \BOthers {.}}{%
Diaz%
\ \protect \BOthers {.}}{%
{\protect \APACyear {2023}}%
}]{%
10.3389/fspas.2023.1117811}
\APACinsertmetastar {%
10.3389/fspas.2023.1117811}%
\begin{APACrefauthors}%
Diaz, J.%
, Kuhlman, B\BPBI M.%
, Edenhoffer, N\BPBI P.%
, Evans, A\BPBI C.%
, Martin, K\BPBI A.%
, Guida, P.%
\BDBL {}Porada, C\BPBI D.%
\end{APACrefauthors}%
\unskip\
\newblock
\APACrefYearMonthDay{2023}{}{}.
\newblock
{\BBOQ}\APACrefatitle {Immediate effects of acute Mars mission equivalent doses
  of SEP and GCR radiation on the murine gastrointestinal system-protective
  effects of curcumin-loaded nanolipoprotein particles (cNLPs)} {Immediate
  effects of acute mars mission equivalent doses of sep and gcr radiation on
  the murine gastrointestinal system-protective effects of curcumin-loaded
  nanolipoprotein particles (cnlps)}.{\BBCQ}
\newblock
\APACjournalVolNumPages{Frontiers in Astronomy and Space Sciences}{10}{}{}.
\newblock
\begin{APACrefURL}
  \url{https://www.frontiersin.org/journals/astronomy-and-space-sciences/articles/10.3389/fspas.2023.1117811}
  \end{APACrefURL}
\newblock
\begin{APACrefDOI} \doi{10.3389/fspas.2023.1117811} \end{APACrefDOI}
\PrintBackRefs{\CurrentBib}

\bibitem [\protect \citeauthoryear {%
Dobynde%
, Shprits%
, Drozdov%
, Hoffman%
\BCBL {}\ \BBA {} Li%
}{%
Dobynde%
\ \protect \BOthers {.}}{%
{\protect \APACyear {2021}}%
}]{%
https://doi.org/10.1029/2021SW002749}
\APACinsertmetastar {%
https://doi.org/10.1029/2021SW002749}%
\begin{APACrefauthors}%
Dobynde, M\BPBI I.%
, Shprits, Y\BPBI Y.%
, Drozdov, A\BPBI Y.%
, Hoffman, J.%
\BCBL {}\ \BBA {} Li, J.%
\end{APACrefauthors}%
\unskip\
\newblock
\APACrefYearMonthDay{2021}{}{}.
\newblock
{\BBOQ}\APACrefatitle {Beating 1 Sievert: Optimal Radiation Shielding of
  Astronauts on a Mission to Mars} {Beating 1 sievert: Optimal radiation
  shielding of astronauts on a mission to mars}.{\BBCQ}
\newblock
\APACjournalVolNumPages{Space Weather}{19}{9}{e2021SW002749}.
\newblock
\begin{APACrefURL}
  \url{https://agupubs.onlinelibrary.wiley.com/doi/abs/10.1029/2021SW002749}
  \end{APACrefURL}
\newblock
\APACrefnote{e2021SW002749 2021SW002749}
\newblock
\begin{APACrefDOI} \doi{https://doi.org/10.1029/2021SW002749} \end{APACrefDOI}
\PrintBackRefs{\CurrentBib}

\bibitem [\protect \citeauthoryear {%
Drout%
\ \protect \BOthers {.}}{%
Drout%
\ \protect \BOthers {.}}{%
{\protect \APACyear {2017}}%
}]{%
doi:10.1126/science.aaq0049}
\APACinsertmetastar {%
doi:10.1126/science.aaq0049}%
\begin{APACrefauthors}%
Drout, M\BPBI R.%
, Piro, A\BPBI L.%
, Shappee, B\BPBI J.%
, Kilpatrick, C\BPBI D.%
, Simon, J\BPBI D.%
, Contreras, C.%
\BDBL {}Whitten, D\BPBI D.%
\end{APACrefauthors}%
\unskip\
\newblock
\APACrefYearMonthDay{2017}{}{}.
\newblock
{\BBOQ}\APACrefatitle {Light curves of the neutron star merger GW170817/SSS17a:
  Implications for r-process nucleosynthesis} {Light curves of the neutron star
  merger gw170817/sss17a: Implications for r-process nucleosynthesis}.{\BBCQ}
\newblock
\APACjournalVolNumPages{Science}{358}{6370}{1570-1574}.
\newblock
\begin{APACrefURL}
  \url{https://www.science.org/doi/abs/10.1126/science.aaq0049}
  \end{APACrefURL}
\newblock
\begin{APACrefDOI} \doi{10.1126/science.aaq0049} \end{APACrefDOI}
\PrintBackRefs{\CurrentBib}

\bibitem [\protect \citeauthoryear {%
Dumbovic%
, Heber%
, Vršnak%
, Temmer%
\BCBL {}\ \BBA {} Kirin%
}{%
Dumbovic%
\ \protect \BOthers {.}}{%
{\protect \APACyear {2018}}%
}]{%
Dumbovic_2018}
\APACinsertmetastar {%
Dumbovic_2018}%
\begin{APACrefauthors}%
Dumbovic, M.%
, Heber, B.%
, Vršnak, B.%
, Temmer, M.%
\BCBL {}\ \BBA {} Kirin, A.%
\end{APACrefauthors}%
\unskip\
\newblock
\APACrefYearMonthDay{2018}{jun}{}.
\newblock
{\BBOQ}\APACrefatitle {An Analytical Diffusion–Expansion Model for Forbush
  Decreases Caused by Flux Ropes} {An analytical diffusion–expansion model
  for forbush decreases caused by flux ropes}.{\BBCQ}
\newblock
\APACjournalVolNumPages{The Astrophysical Journal}{860}{1}{71}.
\newblock
\begin{APACrefURL} \url{https://dx.doi.org/10.3847/1538-4357/aac2de}
  \end{APACrefURL}
\newblock
\begin{APACrefDOI} \doi{10.3847/1538-4357/aac2de} \end{APACrefDOI}
\PrintBackRefs{\CurrentBib}

\bibitem [\protect \citeauthoryear {%
Ehresmann%
\ \protect \BOthers {.}}{%
Ehresmann%
\ \protect \BOthers {.}}{%
{\protect \APACyear {2021}}%
}]{%
ehresmann2021natural}
\APACinsertmetastar {%
ehresmann2021natural}%
\begin{APACrefauthors}%
Ehresmann, B.%
, Hassler, D.%
, Zeitlin, C.%
, Guo, J.%
, Wimmer-Schweingruber, R.%
, Khaksari, S.%
\BCBL {}\ \BBA {} Loeffler, S.%
\end{APACrefauthors}%
\unskip\
\newblock
\APACrefYearMonthDay{2021}{}{}.
\newblock
{\BBOQ}\APACrefatitle {Natural radiation shielding on {Mars} measured with the
  {MSL/RAD} instrument} {Natural radiation shielding on {Mars} measured with
  the {MSL/RAD} instrument}.{\BBCQ}
\newblock
\APACjournalVolNumPages{Journal of Geophysical Research:
  Planets}{126}{8}{e2021JE006851}.
\newblock
\begin{APACrefDOI} \doi{https://doi.org/10.1029/2021JE006851} \end{APACrefDOI}
\PrintBackRefs{\CurrentBib}

\bibitem [\protect \citeauthoryear {%
Ehresmann%
\ \protect \BOthers {.}}{%
Ehresmann%
\ \protect \BOthers {.}}{%
{\protect \APACyear {2018}}%
}]{%
https://doi.org/10.1029/2018GL077801}
\APACinsertmetastar {%
https://doi.org/10.1029/2018GL077801}%
\begin{APACrefauthors}%
Ehresmann, B.%
, Hassler, D\BPBI M.%
, Zeitlin, C.%
, Guo, J.%
, Wimmer-Schweingruber, R\BPBI F.%
, Matthiä, D.%
\BDBL {}Reitz, G.%
\end{APACrefauthors}%
\unskip\
\newblock
\APACrefYearMonthDay{2018}{}{}.
\newblock
{\BBOQ}\APACrefatitle {Energetic Particle Radiation Environment Observed by RAD
  on the Surface of Mars During the September 2017 Event} {Energetic particle
  radiation environment observed by rad on the surface of mars during the
  september 2017 event}.{\BBCQ}
\newblock
\APACjournalVolNumPages{Geophysical Research Letters}{45}{11}{5305-5311}.
\newblock
\begin{APACrefURL}
  \url{https://agupubs.onlinelibrary.wiley.com/doi/abs/10.1029/2018GL077801}
  \end{APACrefURL}
\newblock
\begin{APACrefDOI} \doi{https://doi.org/10.1029/2018GL077801} \end{APACrefDOI}
\PrintBackRefs{\CurrentBib}

\bibitem [\protect \citeauthoryear {%
Ehresmann%
\ \protect \BOthers {.}}{%
Ehresmann%
\ \protect \BOthers {.}}{%
{\protect \APACyear {2014}}%
}]{%
https://doi.org/10.1002/2013JE004547}
\APACinsertmetastar {%
https://doi.org/10.1002/2013JE004547}%
\begin{APACrefauthors}%
Ehresmann, B.%
, Zeitlin, C.%
, Hassler, D\BPBI M.%
, Wimmer-Schweingruber, R\BPBI F.%
, Böhm, E.%
, Böttcher, S.%
\BDBL {}Reitz, G.%
\end{APACrefauthors}%
\unskip\
\newblock
\APACrefYearMonthDay{2014}{}{}.
\newblock
{\BBOQ}\APACrefatitle {Charged particle spectra obtained with the Mars Science
  Laboratory Radiation Assessment Detector (MSL/RAD) on the surface of Mars}
  {Charged particle spectra obtained with the mars science laboratory radiation
  assessment detector (msl/rad) on the surface of mars}.{\BBCQ}
\newblock
\APACjournalVolNumPages{Journal of Geophysical Research:
  Planets}{119}{3}{468-479}.
\newblock
\begin{APACrefURL}
  \url{https://agupubs.onlinelibrary.wiley.com/doi/abs/10.1002/2013JE004547}
  \end{APACrefURL}
\newblock
\begin{APACrefDOI} \doi{https://doi.org/10.1002/2013JE004547} \end{APACrefDOI}
\PrintBackRefs{\CurrentBib}

\bibitem [\protect \citeauthoryear {%
Feldman%
\ \protect \BOthers {.}}{%
Feldman%
\ \protect \BOthers {.}}{%
{\protect \APACyear {2002}}%
}]{%
feldman2002global}
\APACinsertmetastar {%
feldman2002global}%
\begin{APACrefauthors}%
Feldman, W.%
, Boynton, W.%
, Tokar, R.%
, Prettyman, T.%
, Gasnault, O.%
, Squyres, S.%
\BDBL {}others%
\end{APACrefauthors}%
\unskip\
\newblock
\APACrefYearMonthDay{2002}{}{}.
\newblock
{\BBOQ}\APACrefatitle {Global distribution of neutrons from Mars: Results from
  Mars Odyssey} {Global distribution of neutrons from mars: Results from mars
  odyssey}.{\BBCQ}
\newblock
\APACjournalVolNumPages{Science}{297}{5578}{75--78}.
\PrintBackRefs{\CurrentBib}

\bibitem [\protect \citeauthoryear {%
Gieseler%
}{%
Gieseler%
}{%
{\protect \APACyear {2018}}%
}]{%
gieseler2018}
\APACinsertmetastar {%
gieseler2018}%
\begin{APACrefauthors}%
Gieseler, J.%
\end{APACrefauthors}%
\unskip\
\newblock
\APACrefYear{2018}.
\unskip\
\newblock
\APACrefbtitle {Understanding Galactic Cosmic Ray Modulation: Observations and
  Theory} {Understanding galactic cosmic ray modulation: Observations and
  theory}\ \APACtypeAddressSchool {thesis}{}{}.
\PrintBackRefs{\CurrentBib}

\bibitem [\protect \citeauthoryear {%
Gohel%
\ \BBA {} Makwana%
}{%
Gohel%
\ \BBA {} Makwana%
}{%
{\protect \APACyear {2022}}%
}]{%
GOHEL2022110131}
\APACinsertmetastar {%
GOHEL2022110131}%
\begin{APACrefauthors}%
Gohel, A.%
\BCBT {}\ \BBA {} Makwana, R.%
\end{APACrefauthors}%
\unskip\
\newblock
\APACrefYearMonthDay{2022}{}{}.
\newblock
{\BBOQ}\APACrefatitle {Multi-layered shielding materials for high energy space
  radiation} {Multi-layered shielding materials for high energy space
  radiation}.{\BBCQ}
\newblock
\APACjournalVolNumPages{Radiation Physics and Chemistry}{197}{}{110131}.
\newblock
\begin{APACrefURL}
  \url{https://www.sciencedirect.com/science/article/pii/S0969806X22001736}
  \end{APACrefURL}
\newblock
\begin{APACrefDOI} \doi{https://doi.org/10.1016/j.radphyschem.2022.110131}
  \end{APACrefDOI}
\PrintBackRefs{\CurrentBib}

\bibitem [\protect \citeauthoryear {%
Grotzinger%
\ \protect \BOthers {.}}{%
Grotzinger%
\ \protect \BOthers {.}}{%
{\protect \APACyear {2012}}%
}]{%
Grotzinger2012}
\APACinsertmetastar {%
Grotzinger2012}%
\begin{APACrefauthors}%
Grotzinger, J\BPBI P.%
, Crisp, J.%
, Vasavada, A\BPBI R.%
, Anderson, R\BPBI C.%
, Baker, C\BPBI J.%
, Barry, R.%
\BDBL {}Wiens, R\BPBI C.%
\end{APACrefauthors}%
\unskip\
\newblock
\APACrefYearMonthDay{2012}{{\APACmonth{09}}}{}.
\newblock
{\BBOQ}\APACrefatitle {Mars Science Laboratory Mission and Science
  Investigation} {Mars science laboratory mission and science
  investigation}.{\BBCQ}
\newblock
\APACjournalVolNumPages{Space Science Reviews}{170}{1}{5--56}.
\newblock
\begin{APACrefURL} \url{https://doi.org/10.1007/s11214-012-9892-2}
  \end{APACrefURL}
\newblock
\begin{APACrefDOI} \doi{10.1007/s11214-012-9892-2} \end{APACrefDOI}
\PrintBackRefs{\CurrentBib}

\bibitem [\protect \citeauthoryear {%
{Guo}%
\ \protect \BOthers {.}}{%
{Guo}%
\ \protect \BOthers {.}}{%
{\protect \APACyear {2018}}%
}]{%
refId0}
\APACinsertmetastar {%
refId0}%
\begin{APACrefauthors}%
{Guo}%
, {Lillis, Robert}%
, {Wimmer-Schweingruber, Robert F.}%
, {Zeitlin, Cary}%
, {Simonson, Patrick}%
, {Rahmati, Ali}%
\BDBL {}{Böttcher, Stephan}%
\end{APACrefauthors}%
\unskip\
\newblock
\APACrefYearMonthDay{2018}{}{}.
\newblock
{\BBOQ}\APACrefatitle {Measurements of Forbush decreases at Mars: both by MSL
  on ground and by MAVEN in orbit} {Measurements of forbush decreases at mars:
  both by msl on ground and by maven in orbit}.{\BBCQ}
\newblock
\APACjournalVolNumPages{A\&A}{611}{}{A79}.
\newblock
\begin{APACrefURL} \url{https://doi.org/10.1051/0004-6361/201732087}
  \end{APACrefURL}
\newblock
\begin{APACrefDOI} \doi{10.1051/0004-6361/201732087} \end{APACrefDOI}
\PrintBackRefs{\CurrentBib}

\bibitem [\protect \citeauthoryear {%
Guo%
, Khaksarighiri%
\BCBL {}\ \protect \BOthers {.}}{%
Guo%
, Khaksarighiri%
\BCBL {}\ \protect \BOthers {.}}{%
{\protect \APACyear {2021}}%
}]{%
https://doi.org/10.1029/2021GL093912}
\APACinsertmetastar {%
https://doi.org/10.1029/2021GL093912}%
\begin{APACrefauthors}%
Guo, J.%
, Khaksarighiri, S.%
, Wimmer-Schweingruber, R\BPBI F.%
, Hassler, D\BPBI M.%
, Ehresmann, B.%
, Zeitlin, C.%
\BDBL {}Calef, F.%
\end{APACrefauthors}%
\unskip\
\newblock
\APACrefYearMonthDay{2021}{}{}.
\newblock
{\BBOQ}\APACrefatitle {Directionality of the Martian Surface Radiation and
  Derivation of the Upward Albedo Radiation} {Directionality of the martian
  surface radiation and derivation of the upward albedo radiation}.{\BBCQ}
\newblock
\APACjournalVolNumPages{Geophysical Research Letters}{48}{15}{e2021GL093912}.
\newblock
\begin{APACrefURL}
  \url{https://agupubs.onlinelibrary.wiley.com/doi/abs/10.1029/2021GL093912}
  \end{APACrefURL}
\newblock
\APACrefnote{e2021GL093912 2021GL093912}
\newblock
\begin{APACrefDOI} \doi{https://doi.org/10.1029/2021GL093912} \end{APACrefDOI}
\PrintBackRefs{\CurrentBib}

\bibitem [\protect \citeauthoryear {%
Guo%
\ \protect \BOthers {.}}{%
Guo%
\ \protect \BOthers {.}}{%
{\protect \APACyear {2023}}%
}]{%
https://doi.org/10.1029/2023GL103069}
\APACinsertmetastar {%
https://doi.org/10.1029/2023GL103069}%
\begin{APACrefauthors}%
Guo, J.%
, Li, X.%
, Zhang, J.%
, Dobynde, M\BPBI I.%
, Wang, Y.%
, Xu, Z.%
\BDBL {}Zhuang, B.%
\end{APACrefauthors}%
\unskip\
\newblock
\APACrefYearMonthDay{2023}{}{}.
\newblock
{\BBOQ}\APACrefatitle {The First Ground Level Enhancement Seen on Three
  Planetary Surfaces: Earth, Moon, and Mars} {The first ground level
  enhancement seen on three planetary surfaces: Earth, moon, and mars}.{\BBCQ}
\newblock
\APACjournalVolNumPages{Geophysical Research Letters}{50}{15}{e2023GL103069}.
\newblock
\begin{APACrefURL}
  \url{https://agupubs.onlinelibrary.wiley.com/doi/abs/10.1029/2023GL103069}
  \end{APACrefURL}
\newblock
\APACrefnote{e2023GL103069 2023GL103069}
\newblock
\begin{APACrefDOI} \doi{https://doi.org/10.1029/2023GL103069} \end{APACrefDOI}
\PrintBackRefs{\CurrentBib}

\bibitem [\protect \citeauthoryear {%
Guo%
, Slaba%
\BCBL {}\ \protect \BOthers {.}}{%
Guo%
, Slaba%
\BCBL {}\ \protect \BOthers {.}}{%
{\protect \APACyear {2017}}%
}]{%
https://doi.org/10.1002/2016JE005206}
\APACinsertmetastar {%
https://doi.org/10.1002/2016JE005206}%
\begin{APACrefauthors}%
Guo, J.%
, Slaba, T\BPBI C.%
, Zeitlin, C.%
, Wimmer-Schweingruber, R\BPBI F.%
, Badavi, F\BPBI F.%
, Böhm, E.%
\BDBL {}Rafkin, S.%
\end{APACrefauthors}%
\unskip\
\newblock
\APACrefYearMonthDay{2017}{}{}.
\newblock
{\BBOQ}\APACrefatitle {Dependence of the Martian radiation environment on
  atmospheric depth: Modeling and measurement} {Dependence of the martian
  radiation environment on atmospheric depth: Modeling and measurement}.{\BBCQ}
\newblock
\APACjournalVolNumPages{Journal of Geophysical Research:
  Planets}{122}{2}{329-341}.
\newblock
\begin{APACrefURL}
  \url{https://agupubs.onlinelibrary.wiley.com/doi/abs/10.1002/2016JE005206}
  \end{APACrefURL}
\newblock
\begin{APACrefDOI} \doi{https://doi.org/10.1002/2016JE005206} \end{APACrefDOI}
\PrintBackRefs{\CurrentBib}

\bibitem [\protect \citeauthoryear {%
Guo%
, Wimmer-Schweingruber%
, Grande%
, Hannah Lee-Payne%
\BCBL {}\ \BBA {} Matthia%
}{%
Guo%
\ \protect \BOthers {.}}{%
{\protect \APACyear {2019}}%
}]{%
Guo_2019_Cutoff}
\APACinsertmetastar {%
Guo_2019_Cutoff}%
\begin{APACrefauthors}%
Guo, J.%
, Wimmer-Schweingruber, R\BPBI F.%
, Grande, M.%
, Hannah Lee-Payne, Z.%
\BCBL {}\ \BBA {} Matthia, D.%
\end{APACrefauthors}%
\unskip\
\newblock
\APACrefYearMonthDay{2019}{}{}.
\newblock
{\BBOQ}\APACrefatitle {Ready functions for calculating the Martian radiation
  environment} {Ready functions for calculating the martian radiation
  environment}.{\BBCQ}
\newblock
\APACjournalVolNumPages{Journal of Space Weather and Space Climate}{9}{}{A7}.
\newblock
\begin{APACrefURL} \url{http://dx.doi.org/10.1051/swsc/2019004}
  \end{APACrefURL}
\newblock
\begin{APACrefDOI} \doi{10.1051/swsc/2019004} \end{APACrefDOI}
\PrintBackRefs{\CurrentBib}

\bibitem [\protect \citeauthoryear {%
{Guo}%
\ \protect \BOthers {.}}{%
{Guo}%
\ \protect \BOthers {.}}{%
{\protect \APACyear {2021}}%
}]{%
2021A&ARv..29....8G}
\APACinsertmetastar {%
2021A&ARv..29....8G}%
\begin{APACrefauthors}%
{Guo}, J.%
, {Zeitlin}, C.%
, {Wimmer-Schweingruber}, R\BPBI F.%
, {Hassler}, D\BPBI M.%
, {Ehresmann}, B.%
, {Rafkin}, S.%
\BDBL {}{Wang}, Y.%
\end{APACrefauthors}%
\unskip\
\newblock
\APACrefYearMonthDay{2021}{{\APACmonth{12}}}{}.
\newblock
{\BBOQ}\APACrefatitle {{Radiation environment for future human exploration on
  the surface of Mars: the current understanding based on MSL/RAD dose
  measurements}} {{Radiation environment for future human exploration on the
  surface of Mars: the current understanding based on MSL/RAD dose
  measurements}}.{\BBCQ}
\newblock
\APACjournalVolNumPages{aapr}{29}{1}{8}.
\newblock
\begin{APACrefDOI} \doi{10.1007/s00159-021-00136-5} \end{APACrefDOI}
\PrintBackRefs{\CurrentBib}

\bibitem [\protect \citeauthoryear {%
Guo%
, Zeitlin%
\BCBL {}\ \protect \BOthers {.}}{%
Guo%
, Zeitlin%
\BCBL {}\ \protect \BOthers {.}}{%
{\protect \APACyear {2017}}%
}]{%
GUO201712}
\APACinsertmetastar {%
GUO201712}%
\begin{APACrefauthors}%
Guo, J.%
, Zeitlin, C.%
, Wimmer-Schweingruber, R.%
, Hassler, D\BPBI M.%
, Köhler, J.%
, Ehresmann, B.%
\BDBL {}Brinza, D\BPBI E.%
\end{APACrefauthors}%
\unskip\
\newblock
\APACrefYearMonthDay{2017}{}{}.
\newblock
{\BBOQ}\APACrefatitle {Measurements of the neutral particle spectra on Mars by
  MSL/RAD from 2015-11-15 to 2016-01-15} {Measurements of the neutral particle
  spectra on mars by msl/rad from 2015-11-15 to 2016-01-15}.{\BBCQ}
\newblock
\APACjournalVolNumPages{Life Sciences in Space Research}{14}{}{12-17}.
\newblock
\begin{APACrefURL}
  \url{https://www.sciencedirect.com/science/article/pii/S2214552417300093}
  \end{APACrefURL}
\newblock
\APACrefnote{Radiation on the Martian Surface: Model Comparisons with Data from
  the Radiation Assessment Detector on the Mars Science Laboratory (MSL/RAD):
  Results from the 1st Mars Space Radiation Modeling Workshop}
\newblock
\begin{APACrefDOI} \doi{https://doi.org/10.1016/j.lssr.2017.06.001}
  \end{APACrefDOI}
\PrintBackRefs{\CurrentBib}

\bibitem [\protect \citeauthoryear {%
Guo%
\ \protect \BOthers {.}}{%
Guo%
\ \protect \BOthers {.}}{%
{\protect \APACyear {2015}}%
}]{%
guo2015modeling}
\APACinsertmetastar {%
guo2015modeling}%
\begin{APACrefauthors}%
Guo, J.%
, Zeitlin, C.%
, Wimmer-Schweingruber, R\BPBI F.%
, Rafkin, S.%
, Hassler, D\BPBI M.%
, Posner, A.%
\BDBL {}others%
\end{APACrefauthors}%
\unskip\
\newblock
\APACrefYearMonthDay{2015}{}{}.
\newblock
{\BBOQ}\APACrefatitle {Modeling the Variations of Dose Rate Measured by {RAD}
  during the First {MSL} {M}artian Year: 2012--2014} {Modeling the variations
  of dose rate measured by {RAD} during the first {MSL} {M}artian year:
  2012--2014}.{\BBCQ}
\newblock
\APACjournalVolNumPages{The Astrophysical Journal}{810}{1}{24}.
\newblock
\begin{APACrefDOI} \doi{10.1088/0004-637X/810/1/24} \end{APACrefDOI}
\PrintBackRefs{\CurrentBib}

\bibitem [\protect \citeauthoryear {%
Guo%
, Zeitlin%
\BCBL {}\ \protect \BOthers {.}}{%
Guo%
, Zeitlin%
\BCBL {}\ \protect \BOthers {.}}{%
{\protect \APACyear {2021}}%
}]{%
guo2021radiation}
\APACinsertmetastar {%
guo2021radiation}%
\begin{APACrefauthors}%
Guo, J.%
, Zeitlin, R\BPBI F., Cary Wimmer-Schweingruber%
, Hassler, D\BPBI M.%
, Ehresmann, B.%
, Rafkin, S.%
, Freiherr~von Forstner, J\BPBI L.%
\BDBL {}Wang, Y.%
\end{APACrefauthors}%
\unskip\
\newblock
\APACrefYearMonthDay{2021}{}{}.
\newblock
{\BBOQ}\APACrefatitle {Radiation environment for future human exploration on
  the surface of {Mars}: the current understanding based on {MSL/RAD} dose
  measurements} {Radiation environment for future human exploration on the
  surface of {Mars}: the current understanding based on {MSL/RAD} dose
  measurements}.{\BBCQ}
\newblock
\APACjournalVolNumPages{The Astronomy and Astrophysics Review}{29}{1}{1--81}.
\newblock
\begin{APACrefDOI} \doi{https://doi.org/10.1007/s00159-021-00136-5}
  \end{APACrefDOI}
\PrintBackRefs{\CurrentBib}

\bibitem [\protect \citeauthoryear {%
Gómez-Elvira%
\ \protect \BOthers {.}}{%
Gómez-Elvira%
\ \protect \BOthers {.}}{%
{\protect \APACyear {2012}}%
}]{%
Gomez-Elvira2012}
\APACinsertmetastar {%
Gomez-Elvira2012}%
\begin{APACrefauthors}%
Gómez-Elvira, J.%
, Armiens, C.%
, Castañer, L.%
, Domínguez, M.%
, Genzer, M.%
, Gómez, F.%
\BDBL {}Martín-Torres, J.%
\end{APACrefauthors}%
\unskip\
\newblock
\APACrefYearMonthDay{2012}{}{}.
\newblock
{\BBOQ}\APACrefatitle {{REMS}: The Environmental Sensor Suite for the Mars
  Science Laboratory Rover} {{REMS}: The environmental sensor suite for the
  mars science laboratory rover}.{\BBCQ}
\newblock
\APACjournalVolNumPages{Space Science Reviews}{170}{1}{583--640}.
\newblock
\begin{APACrefURL} \url{https://doi.org/10.1007/s11214-012-9921-1}
  \end{APACrefURL}
\newblock
\begin{APACrefDOI} \doi{10.1007/s11214-012-9921-1} \end{APACrefDOI}
\PrintBackRefs{\CurrentBib}

\bibitem [\protect \citeauthoryear {%
HandWiki%
}{%
HandWiki%
}{%
{\protect \APACyear {2024}}%
}]{%
HandWiki2024}
\APACinsertmetastar {%
HandWiki2024}%
\begin{APACrefauthors}%
HandWiki.%
\end{APACrefauthors}%
\unskip\
\newblock
\APACrefYearMonthDay{2024}{}{}.
\newblock
\APACrefbtitle {Solar Storm of August 1972.} {Solar storm of august 1972.}
\newblock
\APAChowpublished {\url{https://encyclopedia.pub/entry/36949}}.
\newblock
\APACaddressPublisher{}{Encyclopedia}.
\newblock
\APACrefnote{Accessed: 2024-11-20}
\PrintBackRefs{\CurrentBib}

\bibitem [\protect \citeauthoryear {%
Harrison%
\ \protect \BOthers {.}}{%
Harrison%
\ \protect \BOthers {.}}{%
{\protect \APACyear {2021}}%
}]{%
harrison2021icrp}
\APACinsertmetastar {%
harrison2021icrp}%
\begin{APACrefauthors}%
Harrison, J.%
, Balonov, M.%
, Bochud, F.%
, Martin, C.%
, Menzel, H.%
, Ortiz-Lopez, P.%
\BDBL {}Wakeford, R.%
\end{APACrefauthors}%
\unskip\
\newblock
\APACrefYearMonthDay{2021}{}{}.
\newblock
{\BBOQ}\APACrefatitle {ICRP publication 147: use of dose quantities in
  radiological protection} {Icrp publication 147: use of dose quantities in
  radiological protection}.{\BBCQ}
\newblock
\APACjournalVolNumPages{Annals of the ICRP}{50}{1}{9--82}.
\newblock
\begin{APACrefDOI} \doi{https://doi.org/10.1177/0146645320911864}
  \end{APACrefDOI}
\PrintBackRefs{\CurrentBib}

\bibitem [\protect \citeauthoryear {%
Hassler%
\ \protect \BOthers {.}}{%
Hassler%
\ \protect \BOthers {.}}{%
{\protect \APACyear {2012}}%
}]{%
hassler2012radiation}
\APACinsertmetastar {%
hassler2012radiation}%
\begin{APACrefauthors}%
Hassler, D\BPBI M.%
, Zeitlin, C.%
, Wimmer-Schweingruber, R.%
, B{\"o}ttcher, S.%
, Martin, C.%
, Andrews, J.%
\BDBL {}others%
\end{APACrefauthors}%
\unskip\
\newblock
\APACrefYearMonthDay{2012}{}{}.
\newblock
{\BBOQ}\APACrefatitle {The {Radiation} {Assessment} {Detector} {(RAD)}
  investigation} {The {Radiation} {Assessment} {Detector} {(RAD)}
  investigation}.{\BBCQ}
\newblock
\APACjournalVolNumPages{Space Science Reviews}{170}{1-4}{503--558}.
\newblock
\begin{APACrefDOI} \doi{10.1007/s11214-012-9913-1} \end{APACrefDOI}
\PrintBackRefs{\CurrentBib}

\bibitem [\protect \citeauthoryear {%
Kallenrode%
}{%
Kallenrode%
}{%
{\protect \APACyear {1993}}%
}]{%
kallenrode1993neutral}
\APACinsertmetastar {%
kallenrode1993neutral}%
\begin{APACrefauthors}%
Kallenrode, M\BHBI B.%
\end{APACrefauthors}%
\unskip\
\newblock
\APACrefYearMonthDay{1993}{}{}.
\newblock
{\BBOQ}\APACrefatitle {Neutral lines and azimuthal “transport” of solar
  energetic particles} {Neutral lines and azimuthal “transport” of solar
  energetic particles}.{\BBCQ}
\newblock
\APACjournalVolNumPages{Journal of Geophysical Research: Space
  Physics}{98}{A4}{5573--5591}.
\PrintBackRefs{\CurrentBib}

\bibitem [\protect \citeauthoryear {%
Khaksarighiri%
, Guo%
, Wimmer-Schweingruber%
\BCBL {}\ \BBA {} Narici%
}{%
Khaksarighiri%
\ \protect \BOthers {.}}{%
{\protect \APACyear {2021}}%
}]{%
khaksarighiri2021easy}
\APACinsertmetastar {%
khaksarighiri2021easy}%
\begin{APACrefauthors}%
Khaksarighiri, S.%
, Guo, J.%
, Wimmer-Schweingruber, R.%
\BCBL {}\ \BBA {} Narici, L.%
\end{APACrefauthors}%
\unskip\
\newblock
\APACrefYearMonthDay{2021}{}{}.
\newblock
{\BBOQ}\APACrefatitle {An easy-to-use function to assess deep space radiation
  in human brains} {An easy-to-use function to assess deep space radiation in
  human brains}.{\BBCQ}
\newblock
\APACjournalVolNumPages{Scientific reports}{11}{1}{11687}.
\newblock
\begin{APACrefDOI} \doi{https://doi.org/10.1038/s41598-021-90695-5}
  \end{APACrefDOI}
\PrintBackRefs{\CurrentBib}

\bibitem [\protect \citeauthoryear {%
Khaksarighiri%
, Guo%
, Wimmer-Schweingruber%
, Narici%
\BCBL {}\ \BBA {} Lohf%
}{%
Khaksarighiri%
\ \protect \BOthers {.}}{%
{\protect \APACyear {2020}}%
}]{%
khaksarighiri2020calculation}
\APACinsertmetastar {%
khaksarighiri2020calculation}%
\begin{APACrefauthors}%
Khaksarighiri, S.%
, Guo, J.%
, Wimmer-Schweingruber, R.%
, Narici, L.%
\BCBL {}\ \BBA {} Lohf, H.%
\end{APACrefauthors}%
\unskip\
\newblock
\APACrefYearMonthDay{2020}{}{}.
\newblock
{\BBOQ}\APACrefatitle {Calculation of dose distribution in a realistic brain
  structure and the indication of space radiation influence on human brains}
  {Calculation of dose distribution in a realistic brain structure and the
  indication of space radiation influence on human brains}.{\BBCQ}
\newblock
\APACjournalVolNumPages{Life Sciences in Space Research}{27}{}{33--48}.
\newblock
\begin{APACrefDOI} \doi{https://doi.org/10.1016/j.lssr.2020.07.003}
  \end{APACrefDOI}
\PrintBackRefs{\CurrentBib}

\bibitem [\protect \citeauthoryear {%
Khaksarighiri%
\ \protect \BOthers {.}}{%
Khaksarighiri%
\ \protect \BOthers {.}}{%
{\protect \APACyear {2023}}%
}]{%
khaksarighiri2023zenith}
\APACinsertmetastar {%
khaksarighiri2023zenith}%
\begin{APACrefauthors}%
Khaksarighiri, S.%
, Guo, J.%
, Wimmer-Schweingruber, R\BPBI F.%
, L{\"o}ffler, S.%
, Ehresmann, B.%
, Matthi{\"a}, D.%
\BDBL {}Berger, T.%
\end{APACrefauthors}%
\unskip\
\newblock
\APACrefYearMonthDay{2023}{}{}.
\newblock
{\BBOQ}\APACrefatitle {The Zenith-Angle Dependence of the Downward Radiation
  Dose Rate on the Martian Surface: Modeling Versus MSL/RAD Measurement} {The
  zenith-angle dependence of the downward radiation dose rate on the martian
  surface: Modeling versus msl/rad measurement}.{\BBCQ}
\newblock
\APACjournalVolNumPages{Journal of Geophysical Research:
  Planets}{128}{4}{e2022JE007644}.
\newblock
\begin{APACrefDOI} \doi{https://doi.org/10.1029/2022JE007644} \end{APACrefDOI}
\PrintBackRefs{\CurrentBib}

\bibitem [\protect \citeauthoryear {%
Köhler%
\ \protect \BOthers {.}}{%
Köhler%
\ \protect \BOthers {.}}{%
{\protect \APACyear {2014}}%
}]{%
https://doi.org/10.1002/2013JE004539}
\APACinsertmetastar {%
https://doi.org/10.1002/2013JE004539}%
\begin{APACrefauthors}%
Köhler, J.%
, Zeitlin, C.%
, Ehresmann, B.%
, Wimmer-Schweingruber, R\BPBI F.%
, Hassler, D\BPBI M.%
, Reitz, G.%
\BDBL {}Kortmann, O.%
\end{APACrefauthors}%
\unskip\
\newblock
\APACrefYearMonthDay{2014}{}{}.
\newblock
{\BBOQ}\APACrefatitle {Measurements of the neutron spectrum on the Martian
  surface with MSL/RAD} {Measurements of the neutron spectrum on the martian
  surface with msl/rad}.{\BBCQ}
\newblock
\APACjournalVolNumPages{Journal of Geophysical Research:
  Planets}{119}{3}{594-603}.
\newblock
\begin{APACrefURL}
  \url{https://agupubs.onlinelibrary.wiley.com/doi/abs/10.1002/2013JE004539}
  \end{APACrefURL}
\newblock
\begin{APACrefDOI} \doi{https://doi.org/10.1002/2013JE004539} \end{APACrefDOI}
\PrintBackRefs{\CurrentBib}

\bibitem [\protect \citeauthoryear {%
Laurenza%
, Alberti%
\BCBL {}\ \BBA {} Cliver%
}{%
Laurenza%
\ \protect \BOthers {.}}{%
{\protect \APACyear {2018}}%
}]{%
Laurenza_2018}
\APACinsertmetastar {%
Laurenza_2018}%
\begin{APACrefauthors}%
Laurenza, M.%
, Alberti, T.%
\BCBL {}\ \BBA {} Cliver, E\BPBI W.%
\end{APACrefauthors}%
\unskip\
\newblock
\APACrefYearMonthDay{2018}{apr}{}.
\newblock
{\BBOQ}\APACrefatitle {A Short-term ESPERTA-based Forecast Tool for
  Moderate-to-extreme Solar Proton Events} {A short-term esperta-based forecast
  tool for moderate-to-extreme solar proton events}.{\BBCQ}
\newblock
\APACjournalVolNumPages{The Astrophysical Journal}{857}{2}{107}.
\newblock
\begin{APACrefURL} \url{https://dx.doi.org/10.3847/1538-4357/aab712}
  \end{APACrefURL}
\newblock
\begin{APACrefDOI} \doi{10.3847/1538-4357/aab712} \end{APACrefDOI}
\PrintBackRefs{\CurrentBib}

\bibitem [\protect \citeauthoryear {%
Loewe%
}{%
Loewe%
}{%
{\protect \APACyear {2025}}%
}]{%
loewe_2025_14801849}
\APACinsertmetastar {%
loewe_2025_14801849}%
\begin{APACrefauthors}%
Loewe, J\BPBI L.%
\end{APACrefauthors}%
\unskip\
\newblock
\APACrefYearMonthDay{2025}{{\APACmonth{02}}}{}.
\newblock
\APACrefbtitle {Nowcasting Solar Energetic Particle Events for Mars Missions.}
  {Nowcasting solar energetic particle events for mars missions.}
\newblock
\APACaddressPublisher{}{Zenodo}.
\newblock
\begin{APACrefURL} \url{https://doi.org/10.5281/zenodo.14801849}
  \end{APACrefURL}
\newblock
\begin{APACrefDOI} \doi{10.5281/zenodo.14801849} \end{APACrefDOI}
\PrintBackRefs{\CurrentBib}

\bibitem [\protect \citeauthoryear {%
{Mao}%
, {Pecaut}%
\BCBL {}\ \BBA {} {Gridley}%
}{%
{Mao}%
\ \protect \BOthers {.}}{%
{\protect \APACyear {2021}}%
}]{%
2021hbba.book..263M}
\APACinsertmetastar {%
2021hbba.book..263M}%
\begin{APACrefauthors}%
{Mao}, X\BPBI W.%
, {Pecaut}, M\BPBI J.%
\BCBL {}\ \BBA {} {Gridley}, D\BPBI S.%
\end{APACrefauthors}%
\unskip\
\newblock
\APACrefYearMonthDay{2021}{}{}.
\newblock
{\BBOQ}\APACrefatitle {{Acute Risks of Space Radiation}} {{Acute Risks of Space
  Radiation}}.{\BBCQ}
\newblock
\BIn{} L\BPBI R.~{Young}\ \BBA {} J\BPBI P.~{Sutton}\ (\BEDS), \APACrefbtitle
  {Handbook of Bioastronautics} {Handbook of bioastronautics}\ (\BPG~263-276).
\newblock
\begin{APACrefDOI} \doi{10.1007/978-3-319-12191-8_27} \end{APACrefDOI}
\PrintBackRefs{\CurrentBib}

\bibitem [\protect \citeauthoryear {%
Martinez~Sierra%
\ \protect \BOthers {.}}{%
Martinez~Sierra%
\ \protect \BOthers {.}}{%
{\protect \APACyear {2023}}%
}]{%
https://doi.org/10.1029/2022SW003344}
\APACinsertmetastar {%
https://doi.org/10.1029/2022SW003344}%
\begin{APACrefauthors}%
Martinez~Sierra, L\BPBI M.%
, Jun, I.%
, Ehresmann, B.%
, Zeitlin, C.%
, Guo, J.%
, Litvak, M.%
\BDBL {}Loffler, S.%
\end{APACrefauthors}%
\unskip\
\newblock
\APACrefYearMonthDay{2023}{}{}.
\newblock
{\BBOQ}\APACrefatitle {Unfolding the Neutron Flux Spectrum on the Surface of
  Mars Using the MSL-RAD and Odyssey-HEND Data} {Unfolding the neutron flux
  spectrum on the surface of mars using the msl-rad and odyssey-hend
  data}.{\BBCQ}
\newblock
\APACjournalVolNumPages{Space Weather}{21}{8}{e2022SW003344}.
\newblock
\begin{APACrefURL}
  \url{https://agupubs.onlinelibrary.wiley.com/doi/abs/10.1029/2022SW003344}
  \end{APACrefURL}
\newblock
\APACrefnote{e2022SW003344 2022SW003344}
\newblock
\begin{APACrefDOI} \doi{https://doi.org/10.1029/2022SW003344} \end{APACrefDOI}
\PrintBackRefs{\CurrentBib}

\bibitem [\protect \citeauthoryear {%
Mewaldt%
}{%
Mewaldt%
}{%
{\protect \APACyear {1994}}%
}]{%
Mewaldt1994}
\APACinsertmetastar {%
Mewaldt1994}%
\begin{APACrefauthors}%
Mewaldt, R\BPBI A.%
\end{APACrefauthors}%
\unskip\
\newblock
\APACrefYearMonthDay{1994}{Oct}{}.
\newblock
{\BBOQ}\APACrefatitle {Galactic cosmic ray composition and energy spectra}
  {Galactic cosmic ray composition and energy spectra}.{\BBCQ}
\newblock
\APACjournalVolNumPages{Adv Space Res}{14}{10}{737--747}.
\newblock
\begin{APACrefDOI} \doi{10.1016/0273-1177(94)90536-3} \end{APACrefDOI}
\PrintBackRefs{\CurrentBib}

\bibitem [\protect \citeauthoryear {%
Montesinos%
\ \protect \BOthers {.}}{%
Montesinos%
\ \protect \BOthers {.}}{%
{\protect \APACyear {2021}}%
}]{%
Montesinos2021}
\APACinsertmetastar {%
Montesinos2021}%
\begin{APACrefauthors}%
Montesinos, C\BPBI A.%
, Khalid, R.%
, Cristea, O.%
, Greenberger, J\BPBI S.%
, Epperly, M\BPBI W.%
, Lemon, J\BPBI A.%
\BDBL {}Jones, T\BPBI A.%
\end{APACrefauthors}%
\unskip\
\newblock
\APACrefYearMonthDay{2021}{August}{}.
\newblock
{\BBOQ}\APACrefatitle {Space Radiation Protection Countermeasures in
  Microgravity and Planetary Exploration} {Space radiation protection
  countermeasures in microgravity and planetary exploration}.{\BBCQ}
\newblock
\APACjournalVolNumPages{Life}{11}{8}{829}.
\newblock
\begin{APACrefURL} \url{https://www.mdpi.com/2075-1729/11/8/829}
  \end{APACrefURL}
\newblock
\begin{APACrefDOI} \doi{10.3390/life11080829} \end{APACrefDOI}
\PrintBackRefs{\CurrentBib}

\bibitem [\protect \citeauthoryear {%
Moreno-Villanueva%
, Wong%
, Lu%
, Zhang%
\BCBL {}\ \BBA {} Wu%
}{%
Moreno-Villanueva%
\ \protect \BOthers {.}}{%
{\protect \APACyear {2017}}%
}]{%
Moreno-Villanueva2017}
\APACinsertmetastar {%
Moreno-Villanueva2017}%
\begin{APACrefauthors}%
Moreno-Villanueva, M.%
, Wong, M.%
, Lu, T.%
, Zhang, Y.%
\BCBL {}\ \BBA {} Wu, H.%
\end{APACrefauthors}%
\unskip\
\newblock
\APACrefYearMonthDay{2017}{}{}.
\newblock
{\BBOQ}\APACrefatitle {Interplay of space radiation and microgravity in DNA
  damage and DNA damage response} {Interplay of space radiation and
  microgravity in dna damage and dna damage response}.{\BBCQ}
\newblock
\APACjournalVolNumPages{npj Microgravity}{3}{1}{14}.
\newblock
\begin{APACrefURL} \url{https://doi.org/10.1038/s41526-017-0019-7}
  \end{APACrefURL}
\newblock
\begin{APACrefDOI} \doi{10.1038/s41526-017-0019-7} \end{APACrefDOI}
\PrintBackRefs{\CurrentBib}

\bibitem [\protect \citeauthoryear {%
on Radiological Protection~(ICRP)%
}{%
on Radiological Protection~(ICRP)%
}{%
{\protect \APACyear {2007}}%
}]{%
ICRP2007_new}
\APACinsertmetastar {%
ICRP2007_new}%
\begin{APACrefauthors}%
on Radiological Protection~(ICRP), I\BPBI C.%
\end{APACrefauthors}%
\unskip\
\newblock
\APACrefYearMonthDay{2007}{}{}.
\newblock
{\BBOQ}\APACrefatitle {The 2007 Recommendations of the International Commission
  on Radiological Protection} {The 2007 recommendations of the international
  commission on radiological protection}.{\BBCQ}
\newblock
\APACjournalVolNumPages{Annals of the ICRP}{37}{2-4}{}.
\PrintBackRefs{\CurrentBib}

\bibitem [\protect \citeauthoryear {%
Parsons%
\ \BBA {} Townsend%
}{%
Parsons%
\ \BBA {} Townsend%
}{%
{\protect \APACyear {2000}}%
}]{%
Parsons2000}
\APACinsertmetastar {%
Parsons2000}%
\begin{APACrefauthors}%
Parsons, J\BPBI L.%
\BCBT {}\ \BBA {} Townsend, L\BPBI W.%
\end{APACrefauthors}%
\unskip\
\newblock
\APACrefYearMonthDay{2000}{Jun}{}.
\newblock
{\BBOQ}\APACrefatitle {Interplanetary Crew Dose Rates for the August 1972 Solar
  Particle Event} {Interplanetary crew dose rates for the august 1972 solar
  particle event}.{\BBCQ}
\newblock
\APACjournalVolNumPages{Radiation Research}{153}{6}{729--733}.
\newblock
\APACrefnote{Research Support, U.S. Gov't, Non-P.H.S.}
\newblock
\begin{APACrefDOI} \doi{10.1667/0033-7587(2000)153[0729:icdrft]2.0.co;2}
  \end{APACrefDOI}
\PrintBackRefs{\CurrentBib}

\bibitem [\protect \citeauthoryear {%
{Peterson}%
\ \protect \BOthers {.}}{%
{Peterson}%
\ \protect \BOthers {.}}{%
{\protect \APACyear {2013}}%
}]{%
Peterson2013}
\APACinsertmetastar {%
Peterson2013}%
\begin{APACrefauthors}%
{Peterson}, J.%
, Rafkin, S.%
, Zeitlin, C.%
, Ehresmann, B.%
, Weigle, E.%
, S., Jeffers%
\BCBL {}\ \BBA {} Hassler, D\BPBI M\BPBI o.%
\end{APACrefauthors}%
\unskip\
\newblock
\APACrefYearMonthDay{2013}{}{}.
\newblock
\APACrefbtitle {{MSL} {M}ARS {R}ADIATION {A}SSESSMENT {D}ETECTOR {RDR V1.0,
  MSL-M-RAD-3-RDR-V1.0}.} {{MSL} {M}ars {R}adiation {A}ssessment {D}etector
  {RDR V1.0, MSL-M-RAD-3-RDR-V1.0}.}
\newblock
\APACrefnote{NASA Planetary Data System}
\newblock
\begin{APACrefDOI} \doi{10.17189/1519761} \end{APACrefDOI}
\PrintBackRefs{\CurrentBib}

\bibitem [\protect \citeauthoryear {%
Posner%
, Rother%
, Heber%
, Mueller-Mellin%
\BCBL {}\ \BBA {} Krause%
}{%
Posner%
\ \protect \BOthers {.}}{%
{\protect \APACyear {2008}}%
}]{%
RELEASE}
\APACinsertmetastar {%
RELEASE}%
\begin{APACrefauthors}%
Posner, A.%
, Rother, O.%
, Heber, B.%
, Mueller-Mellin, R.%
\BCBL {}\ \BBA {} Krause, A.%
\end{APACrefauthors}%
\unskip\
\newblock
\APACrefYearMonthDay{2008}{12}{}.
\newblock
{\BBOQ}\APACrefatitle {The RELativistic Electron Alert System for Exploration
  (RELEASE): Scope, Verification and Validation Status, and Intended Future
  Use} {The relativistic electron alert system for exploration (release):
  Scope, verification and validation status, and intended future use}.{\BBCQ}
\newblock
\APACjournalVolNumPages{AGU Fall Meeting Abstracts}{}{}{}.
\PrintBackRefs{\CurrentBib}

\bibitem [\protect \citeauthoryear {%
Posner%
\ \BBA {} Strauss%
}{%
Posner%
\ \BBA {} Strauss%
}{%
{\protect \APACyear {2020}}%
}]{%
https://doi.org/10.1029/2019SW002354}
\APACinsertmetastar {%
https://doi.org/10.1029/2019SW002354}%
\begin{APACrefauthors}%
Posner, A.%
\BCBT {}\ \BBA {} Strauss, R.%
\end{APACrefauthors}%
\unskip\
\newblock
\APACrefYearMonthDay{2020}{}{}.
\newblock
{\BBOQ}\APACrefatitle {Warning Time Analysis From SEP Simulations of a Two-Tier
  REleASE System Applied to Mars Exploration} {Warning time analysis from sep
  simulations of a two-tier release system applied to mars exploration}.{\BBCQ}
\newblock
\APACjournalVolNumPages{Space Weather}{18}{4}{e2019SW002354}.
\newblock
\begin{APACrefURL}
  \url{https://agupubs.onlinelibrary.wiley.com/doi/abs/10.1029/2019SW002354}
  \end{APACrefURL}
\newblock
\APACrefnote{e2019SW002354 2019SW002354}
\newblock
\begin{APACrefDOI} \doi{https://doi.org/10.1029/2019SW002354} \end{APACrefDOI}
\PrintBackRefs{\CurrentBib}

\bibitem [\protect \citeauthoryear {%
{Rafkin}%
\ \protect \BOthers {.}}{%
{Rafkin}%
\ \protect \BOthers {.}}{%
{\protect \APACyear {2013}}%
}]{%
Rafkin2013}
\APACinsertmetastar {%
Rafkin2013}%
\begin{APACrefauthors}%
{Rafkin}, S.%
, Peterson, J.%
, Zeitlin, C.%
, Ehresmann, B.%
, Weigle, E.%
\BCBL {}\ \BBA {} Hassler, D\BPBI M\BPBI o.%
\end{APACrefauthors}%
\unskip\
\newblock
\APACrefYearMonthDay{2013}{}{}.
\newblock
\APACrefbtitle {{MSL} {M}ARS {R}ADIATION {A}SSESSMENT {D}ETECTOR {EDR V1.0,
  MSL-M-RAD-2-EDR-V1.0t}.} {{MSL} {M}ars {R}adiation {A}ssessment {D}etector
  {EDR V1.0, MSL-M-RAD-2-EDR-V1.0t}.}
\newblock
\APACrefnote{NASA Planetary Data System}
\newblock
\begin{APACrefDOI} \doi{10.17189/1519760} \end{APACrefDOI}
\PrintBackRefs{\CurrentBib}

\bibitem [\protect \citeauthoryear {%
Rafkin%
\ \protect \BOthers {.}}{%
Rafkin%
\ \protect \BOthers {.}}{%
{\protect \APACyear {2014}}%
}]{%
rafkin2014diurnal}
\APACinsertmetastar {%
rafkin2014diurnal}%
\begin{APACrefauthors}%
Rafkin, S\BPBI C.%
, Zeitlin, C.%
, Ehresmann, B.%
, Hassler, D.%
, Guo, J.%
, K{\"o}hler, J.%
\BDBL {}others%
\end{APACrefauthors}%
\unskip\
\newblock
\APACrefYearMonthDay{2014}{}{}.
\newblock
{\BBOQ}\APACrefatitle {Diurnal variations of energetic particle radiation at
  the surface of Mars as observed by the Mars Science Laboratory Radiation
  Assessment Detector} {Diurnal variations of energetic particle radiation at
  the surface of mars as observed by the mars science laboratory radiation
  assessment detector}.{\BBCQ}
\newblock
\APACjournalVolNumPages{Journal of Geophysical Research:
  Planets}{119}{6}{1345--1358}.
\PrintBackRefs{\CurrentBib}

\bibitem [\protect \citeauthoryear {%
Reames%
}{%
Reames%
}{%
{\protect \APACyear {1999}}%
}]{%
Reames1999}
\APACinsertmetastar {%
Reames1999}%
\begin{APACrefauthors}%
Reames, D\BPBI V.%
\end{APACrefauthors}%
\unskip\
\newblock
\APACrefYearMonthDay{1999}{10}{}.
\newblock
{\BBOQ}\APACrefatitle {Particle acceleration at the Sun and in the heliosphere}
  {Particle acceleration at the sun and in the heliosphere}.{\BBCQ}
\newblock
\APACjournalVolNumPages{Space Science Reviews}{90}{3}{413-491}.
\newblock
\begin{APACrefURL} \url{https://doi.org/10.1023/A:1005105831781}
  \end{APACrefURL}
\newblock
\begin{APACrefDOI} \doi{10.1023/A:1005105831781} \end{APACrefDOI}
\PrintBackRefs{\CurrentBib}

\bibitem [\protect \citeauthoryear {%
Richardson%
, von Rosenvinge%
, Cane%
, Christian%
\BCBL {}\ \BBA {} Cohen%
}{%
Richardson%
\ \protect \BOthers {.}}{%
{\protect \APACyear {{\protect \bibnodate {}}}}%
}]{%
richardson289labrador}
\APACinsertmetastar {%
richardson289labrador}%
\begin{APACrefauthors}%
Richardson, I.%
, von Rosenvinge, T.%
, Cane, H.%
, Christian, E.%
\BCBL {}\ \BBA {} Cohen, C.%
\end{APACrefauthors}%
\unskip\
\newblock
\APACrefYearMonthDay{{\protect \bibnodate {}}}{}{}.
\newblock
{\BBOQ}\APACrefatitle {Labrador, a. W.,... Stone, EC (2014, apr).> 25 MeV
  Proton Events Observed by the High Energy Telescopes on the STEREO A and B
  Spacecraft and/or at Earth During the First Seven Years of the STEREO
  Mission} {Labrador, a. w.,... stone, ec (2014, apr).> 25 mev proton events
  observed by the high energy telescopes on the stereo a and b spacecraft
  and/or at earth during the first seven years of the stereo mission}.{\BBCQ}
\newblock
\APACjournalVolNumPages{Sol. Phys}{289}{}{3059}.
\PrintBackRefs{\CurrentBib}

\bibitem [\protect \citeauthoryear {%
{R{\"o}stel}%
, {Guo}%
, {Banjac}%
, {Wimmer-Schweingruber}%
\BCBL {}\ \BBA {} {Heber}%
}{%
{R{\"o}stel}%
\ \protect \BOthers {.}}{%
{\protect \APACyear {2020}}%
}]{%
2020JGRE..12506246R}
\APACinsertmetastar {%
2020JGRE..12506246R}%
\begin{APACrefauthors}%
{R{\"o}stel}, L.%
, {Guo}, J.%
, {Banjac}, S.%
, {Wimmer-Schweingruber}, R\BPBI F.%
\BCBL {}\ \BBA {} {Heber}, B.%
\end{APACrefauthors}%
\unskip\
\newblock
\APACrefYearMonthDay{2020}{{\APACmonth{03}}}{}.
\newblock
{\BBOQ}\APACrefatitle {{Subsurface Radiation Environment of Mars and Its
  Implication for Shielding Protection of Future Habitats}} {{Subsurface
  Radiation Environment of Mars and Its Implication for Shielding Protection of
  Future Habitats}}.{\BBCQ}
\newblock
\APACjournalVolNumPages{Journal of Geophysical Research
  (Planets)}{125}{3}{e06246}.
\newblock
\begin{APACrefDOI} \doi{10.1029/2019JE006246} \end{APACrefDOI}
\PrintBackRefs{\CurrentBib}

\bibitem [\protect \citeauthoryear {%
Saganti%
, Cucinotta%
, Wilson%
, Simonsen%
\BCBL {}\ \BBA {} Zeitlin%
}{%
Saganti%
\ \protect \BOthers {.}}{%
{\protect \APACyear {2004}}%
}]{%
saganti2004radiation}
\APACinsertmetastar {%
saganti2004radiation}%
\begin{APACrefauthors}%
Saganti, P\BPBI B.%
, Cucinotta, F\BPBI A.%
, Wilson, J\BPBI W.%
, Simonsen, L\BPBI C.%
\BCBL {}\ \BBA {} Zeitlin, C.%
\end{APACrefauthors}%
\unskip\
\newblock
\APACrefYearMonthDay{2004}{}{}.
\newblock
{\BBOQ}\APACrefatitle {Radiation climate map for analyzing risks to astronauts
  on the Mars surface from galactic cosmic rays} {Radiation climate map for
  analyzing risks to astronauts on the mars surface from galactic cosmic
  rays}.{\BBCQ}
\newblock
\APACjournalVolNumPages{Space Science Reviews}{110}{1}{143--156}.
\newblock
\begin{APACrefURL}
  [{2024-09-01}]\url{https://doi.org/10.1023/B:SPAC.0000021010.20082.1a}
  \end{APACrefURL}
\newblock
\begin{APACrefDOI} \doi{10.1023/B:SPAC.0000021010.20082.1a} \end{APACrefDOI}
\PrintBackRefs{\CurrentBib}

\bibitem [\protect \citeauthoryear {%
Simpson%
}{%
Simpson%
}{%
{\protect \APACyear {1983}}%
}]{%
simpson1983}
\APACinsertmetastar {%
simpson1983}%
\begin{APACrefauthors}%
Simpson, J.%
\end{APACrefauthors}%
\unskip\
\newblock
\APACrefYearMonthDay{1983}{}{}.
\newblock
{\BBOQ}\APACrefatitle {Elemental and isotopic composition of the galactic
  cosmic rays} {Elemental and isotopic composition of the galactic cosmic
  rays}.{\BBCQ}
\newblock
\APACjournalVolNumPages{Annual Review of Nuclear and Particle
  Science}{33}{1}{323--382}.
\newblock
\begin{APACrefDOI} \doi{10.1146/annurev.ns.33.120183.001543} \end{APACrefDOI}
\PrintBackRefs{\CurrentBib}

\bibitem [\protect \citeauthoryear {%
Smart%
}{%
Smart%
}{%
{\protect \APACyear {2017}}%
}]{%
Smart2017}
\APACinsertmetastar {%
Smart2017}%
\begin{APACrefauthors}%
Smart, D.%
\end{APACrefauthors}%
\unskip\
\newblock
\APACrefYearMonthDay{2017}{October}{}.
\newblock
{\BBOQ}\APACrefatitle {Radiation Toxicity in the Central Nervous System:
  Mechanisms and Strategies for Injury Reduction} {Radiation toxicity in the
  central nervous system: Mechanisms and strategies for injury
  reduction}.{\BBCQ}
\newblock
\APACjournalVolNumPages{Seminars in Radiation Oncology}{27}{4}{332--339}.
\newblock
\begin{APACrefDOI} \doi{10.1016/j.semradonc.2017.04.006} \end{APACrefDOI}
\PrintBackRefs{\CurrentBib}

\bibitem [\protect \citeauthoryear {%
Tranquille%
}{%
Tranquille%
}{%
{\protect \APACyear {1994}}%
}]{%
tranquille1994solar}
\APACinsertmetastar {%
tranquille1994solar}%
\begin{APACrefauthors}%
Tranquille, C.%
\end{APACrefauthors}%
\unskip\
\newblock
\APACrefYearMonthDay{1994}{}{}.
\newblock
{\BBOQ}\APACrefatitle {Solar proton events and their effect on space systems}
  {Solar proton events and their effect on space systems}.{\BBCQ}
\newblock
\APACjournalVolNumPages{Radiation Physics and Chemistry}{43}{1-2}{35--45}.
\PrintBackRefs{\CurrentBib}

\bibitem [\protect \citeauthoryear {%
Usoskin%
, Bazilevskaya%
\BCBL {}\ \BBA {} Kovaltsov%
}{%
Usoskin%
\ \protect \BOthers {.}}{%
{\protect \APACyear {2011}}%
}]{%
https://doi.org/10.1029/2010JA016105}
\APACinsertmetastar {%
https://doi.org/10.1029/2010JA016105}%
\begin{APACrefauthors}%
Usoskin, I\BPBI G.%
, Bazilevskaya, G\BPBI A.%
\BCBL {}\ \BBA {} Kovaltsov, G\BPBI A.%
\end{APACrefauthors}%
\unskip\
\newblock
\APACrefYearMonthDay{2011}{}{}.
\newblock
{\BBOQ}\APACrefatitle {Solar modulation parameter for cosmic rays since 1936
  reconstructed from ground-based neutron monitors and ionization chambers}
  {Solar modulation parameter for cosmic rays since 1936 reconstructed from
  ground-based neutron monitors and ionization chambers}.{\BBCQ}
\newblock
\APACjournalVolNumPages{Journal of Geophysical Research: Space
  Physics}{116}{A2}{}.
\newblock
\begin{APACrefURL}
  \url{https://agupubs.onlinelibrary.wiley.com/doi/abs/10.1029/2010JA016105}
  \end{APACrefURL}
\newblock
\begin{APACrefDOI} \doi{https://doi.org/10.1029/2010JA016105} \end{APACrefDOI}
\PrintBackRefs{\CurrentBib}

\bibitem [\protect \citeauthoryear {%
Whitman%
\ \protect \BOthers {.}}{%
Whitman%
\ \protect \BOthers {.}}{%
{\protect \APACyear {2023}}%
}]{%
WHITMAN20235161}
\APACinsertmetastar {%
WHITMAN20235161}%
\begin{APACrefauthors}%
Whitman, K.%
, Egeland, R.%
, Richardson, I\BPBI G.%
, Allison, C.%
, Quinn, P.%
, Barzilla, J.%
\BDBL {}Hosseinzadeh, P.%
\end{APACrefauthors}%
\unskip\
\newblock
\APACrefYearMonthDay{2023}{}{}.
\newblock
{\BBOQ}\APACrefatitle {Review of Solar Energetic Particle Prediction Models}
  {Review of solar energetic particle prediction models}.{\BBCQ}
\newblock
\APACjournalVolNumPages{Advances in Space Research}{72}{12}{5161-5242}.
\newblock
\begin{APACrefURL}
  \url{https://www.sciencedirect.com/science/article/pii/S0273117722007244}
  \end{APACrefURL}
\newblock
\APACrefnote{COSPAR Space Weather Roadmap 2022: Scientific Research and
  Applications}
\newblock
\begin{APACrefDOI} \doi{https://doi.org/10.1016/j.asr.2022.08.006}
  \end{APACrefDOI}
\PrintBackRefs{\CurrentBib}

\bibitem [\protect \citeauthoryear {%
Wiedenbeck%
\ \protect \BOthers {.}}{%
Wiedenbeck%
\ \protect \BOthers {.}}{%
{\protect \APACyear {2007}}%
}]{%
Wiedenbeck2007}
\APACinsertmetastar {%
Wiedenbeck2007}%
\begin{APACrefauthors}%
Wiedenbeck, M\BPBI E.%
, Binns, W\BPBI R.%
, Cummings, A\BPBI C.%
, Davis, A\BPBI J.%
, de Nolfo, G\BPBI A.%
, Israel, M\BPBI H.%
\BDBL {}von Rosenvinge, T\BPBI T.%
\end{APACrefauthors}%
\unskip\
\newblock
\APACrefYearMonthDay{2007}{Jun}{}.
\newblock
{\BBOQ}\APACrefatitle {An Overview of the Origin of Galactic Cosmic Rays as
  Inferred from Observations of Heavy Ion Composition and Spectra} {An overview
  of the origin of galactic cosmic rays as inferred from observations of heavy
  ion composition and spectra}.{\BBCQ}
\newblock
\APACjournalVolNumPages{Space Science Reviews}{130}{1}{415--429}.
\newblock
\begin{APACrefURL} \url{https://doi.org/10.1007/s11214-007-9198-y}
  \end{APACrefURL}
\newblock
\begin{APACrefDOI} \doi{10.1007/s11214-007-9198-y} \end{APACrefDOI}
\PrintBackRefs{\CurrentBib}

\bibitem [\protect \citeauthoryear {%
Williams%
}{%
Williams%
}{%
{\protect \APACyear {2024}}%
}]{%
williams2024mars}
\APACinsertmetastar {%
williams2024mars}%
\begin{APACrefauthors}%
Williams, D\BPBI R.%
\end{APACrefauthors}%
\unskip\
\newblock
\APACrefYearMonthDay{2024}{}{}.
\newblock
\APACrefbtitle {Mars Fact Sheet.} {Mars fact sheet.}
\newblock
\APAChowpublished {NSSDCA, Mail Code 690.1, NASA Goddard Space Flight Center,
  Greenbelt, MD 20771}.
\newblock
\begin{APACrefURL}
  [{2024-03-25}]\url{https://nssdc.gsfc.nasa.gov/planetary/factsheet/marsfact.html}
  \end{APACrefURL}
\newblock
\APACrefnote{Last Updated: 11 January 2024}
\PrintBackRefs{\CurrentBib}

\bibitem [\protect \citeauthoryear {%
Zeitlin%
\ \protect \BOthers {.}}{%
Zeitlin%
\ \protect \BOthers {.}}{%
{\protect \APACyear {2013}}%
}]{%
doi:10.1126/science.1235989}
\APACinsertmetastar {%
doi:10.1126/science.1235989}%
\begin{APACrefauthors}%
Zeitlin, C.%
, Hassler, D\BPBI M.%
, Cucinotta, F\BPBI A.%
, Ehresmann, B.%
, Wimmer-Schweingruber, R\BPBI F.%
, Brinza, D\BPBI E.%
\BDBL {}Reitz, G.%
\end{APACrefauthors}%
\unskip\
\newblock
\APACrefYearMonthDay{2013}{}{}.
\newblock
{\BBOQ}\APACrefatitle {Measurements of Energetic Particle Radiation in Transit
  to Mars on the Mars Science Laboratory} {Measurements of energetic particle
  radiation in transit to mars on the mars science laboratory}.{\BBCQ}
\newblock
\APACjournalVolNumPages{Science}{340}{6136}{1080-1084}.
\newblock
\begin{APACrefURL}
  \url{https://www.science.org/doi/abs/10.1126/science.1235989}
  \end{APACrefURL}
\newblock
\begin{APACrefDOI} \doi{10.1126/science.1235989} \end{APACrefDOI}
\PrintBackRefs{\CurrentBib}

\bibitem [\protect \citeauthoryear {%
Zeitlin%
\ \protect \BOthers {.}}{%
Zeitlin%
\ \protect \BOthers {.}}{%
{\protect \APACyear {2019}}%
}]{%
Zeitlin2019_LET}
\APACinsertmetastar {%
Zeitlin2019_LET}%
\begin{APACrefauthors}%
Zeitlin, C.%
, Hassler, D\BPBI M.%
, Ehresmann, B.%
, Rafkin, S\BPBI C\BPBI R.%
, Guo, J.%
, Wimmer-Schweingruber, R\BPBI F.%
\BDBL {}Matthiä, D.%
\end{APACrefauthors}%
\unskip\
\newblock
\APACrefYearMonthDay{2019}{{\APACmonth{08}}}{}.
\newblock
{\BBOQ}\APACrefatitle {Measurements of radiation quality factor on Mars with
  the Mars Science Laboratory Radiation Assessment Detector} {Measurements of
  radiation quality factor on mars with the mars science laboratory radiation
  assessment detector}.{\BBCQ}
\newblock
\APACjournalVolNumPages{Life Sciences in Space Research}{22}{}{89--97}.
\newblock
\APACrefnote{Epub 2019 Jul 19}
\newblock
\begin{APACrefDOI} \doi{10.1016/j.lssr.2019.07.010} \end{APACrefDOI}
\PrintBackRefs{\CurrentBib}

\bibitem [\protect \citeauthoryear {%
Zeitlin%
\ \protect \BOthers {.}}{%
Zeitlin%
\ \protect \BOthers {.}}{%
{\protect \APACyear {2016}}%
}]{%
Zeitlin2016}
\APACinsertmetastar {%
Zeitlin2016}%
\begin{APACrefauthors}%
Zeitlin, C.%
, Hassler, D\BPBI M.%
, Wimmer-Schweingruber, R\BPBI F.%
, Ehresmann, B.%
, Appel, J.%
, Berger, T.%
\BDBL {}Murakami, T.%
\end{APACrefauthors}%
\unskip\
\newblock
\APACrefYearMonthDay{2016}{}{}.
\newblock
{\BBOQ}\APACrefatitle {Calibration and Characterization of the Radiation
  Assessment Detector (RAD) on Curiosity} {Calibration and characterization of
  the radiation assessment detector (rad) on curiosity}.{\BBCQ}
\newblock
\APACjournalVolNumPages{Space Science Reviews}{201}{1}{201--233}.
\newblock
\begin{APACrefURL} \url{https://doi.org/10.1007/s11214-016-0303-y}
  \end{APACrefURL}
\newblock
\begin{APACrefDOI} \doi{10.1007/s11214-016-0303-y} \end{APACrefDOI}
\PrintBackRefs{\CurrentBib}

\bibitem [\protect \citeauthoryear {%
Zhang%
, Guo%
, Dobynde%
, Wang%
\BCBL {}\ \BBA {} Wimmer-Schweingruber%
}{%
Zhang%
\ \protect \BOthers {.}}{%
{\protect \APACyear {2022}}%
}]{%
https://doi.org/10.1029/2021JE007157}
\APACinsertmetastar {%
https://doi.org/10.1029/2021JE007157}%
\begin{APACrefauthors}%
Zhang, J.%
, Guo, J.%
, Dobynde, M\BPBI I.%
, Wang, Y.%
\BCBL {}\ \BBA {} Wimmer-Schweingruber, R\BPBI F.%
\end{APACrefauthors}%
\unskip\
\newblock
\APACrefYearMonthDay{2022}{}{}.
\newblock
{\BBOQ}\APACrefatitle {From the Top of Martian Olympus to Deep Craters and
  Beneath: Mars Radiation Environment Under Different Atmospheric and Regolith
  Depths} {From the top of martian olympus to deep craters and beneath: Mars
  radiation environment under different atmospheric and regolith
  depths}.{\BBCQ}
\newblock
\APACjournalVolNumPages{Journal of Geophysical Research:
  Planets}{127}{3}{e2021JE007157}.
\newblock
\begin{APACrefURL}
  \url{https://agupubs.onlinelibrary.wiley.com/doi/abs/10.1029/2021JE007157}
  \end{APACrefURL}
\newblock
\APACrefnote{e2021JE007157 2021JE007157}
\newblock
\begin{APACrefDOI} \doi{https://doi.org/10.1029/2021JE007157} \end{APACrefDOI}
\PrintBackRefs{\CurrentBib}

\end{thebibliography}

\end{document}